\documentclass[11pt,a4paper]{article}
\usepackage[T1]{fontenc}
\usepackage{lmodern,microtype}
\usepackage[a4paper,margin=26mm]{geometry}
\usepackage{amsmath,amsthm,needspace,tikz}
\usepackage[hidelinks]{hyperref}
\usepackage{amssymb}
\newtheorem{theorem}{Theorem}
\newtheorem{lemma}[theorem]{Lemma}
\newtheorem{observation}[theorem]{Observation}
\hypersetup{
  pdftitle={Independent set reconfiguration is hard at treewidth four},
  pdfauthor={Hussein Abbas, Haydar Hamie, Amer E. Mouawad, Ali Slim},
  pdfkeywords={independent sets, combinatorial reconfiguration, token sliding, token jumping, treewidth}
}

\title{Independent set reconfiguration is hard\\at treewidth four}
\author{Hussein Abbas \quad Haydar Hamieh \quad Amer E. Mouawad \quad Ali Slim\\[.35em]
  \normalsize American University of Beirut, Lebanon}
\date{}

\begin{document}
\maketitle

\begin{abstract}
\textsc{Independent Set Reconfiguration} asks whether one of two given independent
sets of equal size can be transformed into the other by moving one token
at a time while keeping the occupied vertices independent. In token
sliding, a token moves along an edge. In token jumping, it may move to
any unoccupied vertex. For both rules, the problem is solvable in
polynomial time on trees and is PSPACE-complete on graphs of treewidth
at most \(c\) for some fixed constant \(c\). For each rule, the smallest
treewidth bound that suffices for hardness has long been open.
We narrow this gap by proving that both versions
are NP-hard on graphs of treewidth four.

\smallskip
\noindent\textbf{Keywords.} Independent sets, combinatorial reconfiguration,
token sliding, token jumping, treewidth.
\end{abstract}

\section{Introduction}\label{sec:introduction}

Place two tokens on different leaves of a three-leaf star. Can we move
one of them to the empty leaf while keeping the two occupied vertices
nonadjacent? A token can jump there in one move. If it must slide along
an edge, neither token can move at all. Its first step would take it to
the center, next to the other token.

In \textsc{Independent Set Reconfiguration} (ISR), we ask the same question for two independent sets
\(I_{\mathrm{start}}\) and \(I_{\mathrm{goal}}\) in a graph. Each consists
of \(k\) pairwise nonadjacent vertices. A reconfiguration sequence moves
one token at a time and keeps the occupied vertices independent. Under
token sliding the token must follow an edge. Under token jumping its
destination can be any unoccupied vertex. The tokens are indistinguishable,
\(k\) is part of the input, and the number of moves is unrestricted.

On every tree, token jumping connects any two independent sets of the
same size. This follows from a result
of Kami{\'n}ski, Medvedev, and Milani{\v c} for graphs with no induced
even cycle~\cite{kaminski2012reconfiguration}. Token sliding can
leave tokens stuck, as in the star, but Demaine et al.\ showed that one
can still decide reachability on a tree in linear time~\cite{demaine2015trees}.

How far beyond trees do polynomial-time algorithms extend? Treewidth gives
a way to ask this question one graph class at a time, starting with forests
at width one. Wrochna proved PSPACE-completeness for reconfiguring maximum
independent sets on graphs of bounded bandwidth~\cite{wrochna2018bandwidth}.
Van der Zanden proved PSPACE-completeness for both move rules even on
planar graphs of maximum degree three and bounded
bandwidth~\cite{vanderzanden2015constraint}. Since treewidth is at most
bandwidth, these results imply hardness at some fixed treewidth. They
leave open how small that width can be. Bousquet (and others) conjectured that both
rules admit polynomial-time algorithms at width
two~\cite[Conjecture~68]{bousquet2024configuration}.

\begin{theorem}\label{thm:main}
\textsc{Independent Set Reconfiguration} is NP-hard on graphs of treewidth four under
both token sliding and token jumping. 
\end{theorem}

Using maximum independent sets lets us prove the result for both rules
with one construction. If a token in a maximum independent set could
jump to a nonadjacent vertex, adding that vertex before removing the
token would give a larger independent set. Thus every jump in our
instances is a slide.

\paragraph{Idea of the proof.}
We reduce from \textsc{3-CNF-SAT}~\cite{karp1972reducibility}. We need
each clause to choose a true literal, but connecting clauses directly to
their variables could give the graph large treewidth. Instead, we number
the literal occurrences in clause order and give each its own time to
be ``checked''. All clauses follow a single token on a triangle. We count
its forward moves around the triangle minus its backward moves, so
different visits to the same vertex can mark different occurrences.
Each clause can be checked only at one of its own occurrences.

The starting and target independent sets differ only on one final
five-vertex subgraph. To change its occupied pair, the shared token must
reach the end of the count. Consider its first arrival there and the
last time before that when the count was zero. Each clause has its own
pair of tokens that is forced into different positions at those two
times. To get from the first pair of positions to the second, the clause
must pass a literal check somewhere in between.

Why do all these checks use the same assignment? Each variable has a
token on a triangle of its own. At a check, this token has either made
the same net progress as the shared token, representing false, or made
one full turn more, representing true. While the shared count stays
positive, the graph forbids an intermediate relative position. Since
tokens move one step at a time, a variable cannot switch between the two
values during this part of the reconfiguration sequence. At the time
assigned to an occurrence, its gadget permits a check only if the
literal is true. If the occurrence is left unchecked, its gadget
permits either truth value. Thus one clause cannot pass by using
\(x\) and another by using \(\neg x\),
even if the shared token goes backward. Conversely, given a satisfying
assignment, we set the variable tokens accordingly, move forward, and
check one true literal per clause. At the end we change the final
five-vertex graph. Reversing the earlier moves restores every other
occupied vertex and reaches \(I_{\mathrm{goal}}\).

For a formula with \(s\) literal occurrences, the resulting graph has
\(O(s^2)\) vertices and edges. Its tree decomposition has \(O(s^2)\)
bags, each containing at most five vertices.

\paragraph{Related work.}
Hearn and Demaine studied token sliding using nondeterministic constraint
logic~\cite{hearn2005sliding}, and Ito et al.\ developed a general
framework for reconfiguration problems~\cite{ito2011complexity}. Beyond trees,
polynomial-time algorithms for token sliding are known on claw-free
graphs~\cite{bonsma2014clawfree}, interval graphs~\cite{bonamy2017chordal},
and block graphs~\cite{francis2025block}. The two move rules can also have
different complexity classifications on the same class. On bipartite
graphs, Lokshtanov and Mouawad proved NP-completeness for token jumping
and PSPACE-completeness for token sliding~\cite{lokshtanov2019bipartite}.

A different way to restrict the problem is to bound the number of tokens.
For each fixed treewidth bound, token jumping on an \(n\)-vertex graph
within that bound takes \(f(k)n^{O(1)}\) time, by the result of Lokshtanov et al.\ for
graphs of bounded degeneracy~\cite{lokshtanov2018sparse}. On planar
graphs, Bartier, Bousquet, and Mouawad gave such an algorithm for token
sliding~\cite{bartier2023galactic}. For token jumping, Bousquet and
Cranston showed that a planar instance can be reduced in polynomial time
to an equivalent instance on at most \(42k\) vertices~\cite{bousquet2026linear}.
In our reduction the treewidth is four and the number of tokens grows
with the formula.

General surveys of reconfiguration include those of van den
Heuvel~\cite{vandenheuvel2013change} and Nishimura~\cite{nishimura2018intro}.
Ito~\cite{ito2022invitation} and Nishimura~\cite{nishimura2024reasons}
give further introductions, and Ito and Suzuki compiled a web survey
organized by problem and move rule~\cite{itosuzukiweb}. Mynhardt and
Nasserasr survey reconfiguration of colourings and dominating
sets~\cite{mynhardt2019survey}. For a survey devoted to the parameterized
complexity of reconfiguration problems, see
Bousquet et al.~\cite{bousquet2024survey}.

\section{Preliminaries}\label{sec:preliminaries}

Graphs are finite, simple, and undirected. A \emph{configuration} is an
independent set of \(k\) occupied vertices. An independent set of largest
size is \emph{maximum}.

A \emph{tree decomposition} of \(G\) is a tree whose nodes carry sets of
vertices of \(G\), called \emph{bags}. Every vertex occurs in a bag, the
ends of each edge occur together in a bag, and the bags containing any
one vertex form a connected subtree. The width is the largest bag size
minus one. The treewidth of \(G\) is the minimum such width.

\paragraph{Columns and heights.}
A \emph{column} \(T\) consists of one, two, or three pairwise adjacent
vertices \(t_i\), each with a distinct index \(i \in \{0,1,2\}\).
All indices are interpreted modulo three.
In the gadget proofs, each column contains exactly one token.
Section~\ref{sec:input-sets} establishes that every move in the full
graph preserves the stated token counts.

If the token occupies \(t_i\), we call \(t_i\) the \emph{occupied vertex}
and \(i\) the \emph{current index}. A move from \(t_i\) to \(t_{i+1}\) is called \emph{forward}, while a
move from \(t_i\) to \(t_{i-1}\) is called \emph{backward}.

Along a reconfiguration sequence, the \emph{height} \(h(T)\) is the
number of forward moves minus the number of backward moves within
\(T\), measured from the initial configuration.
If the token initially occupies \(t_a\), its current index is
\(a+h(T) \pmod{3}\). In particular, a complete forward traversal of a triangular column
returns the token to its original vertex while increasing its
height by three.

\paragraph{Paths of columns.}
A \emph{path of columns} of length \(\ell\geq 1\) consists of disjoint
columns \(T_0,\ldots,T_\ell\) in this order. Between consecutive columns,
join vertices with the same index. There are no other edges between
columns. Thus consecutive columns have different current indices. Reading
from \(T_0\) to \(T_\ell\), write \(+\) when the index increases by one
modulo three and \(-\) when it decreases by one.

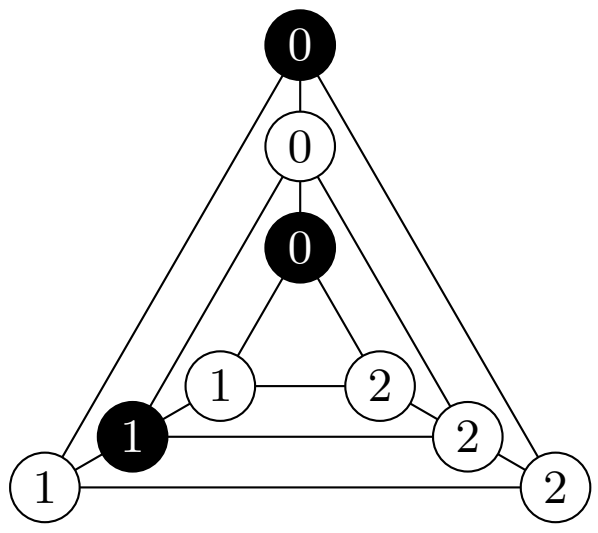
\begin{figure}[!ht]
\centering
\begin{minipage}{.32\linewidth}
\centering
\begin{tikzpicture}[
  vertex/.style={circle,draw,fill=white,inner sep=0pt,minimum size=4.6mm,font=\footnotesize},
  occupied/.style={vertex,fill=black,text=white}]
  \foreach \layer/\radius in {0/1.95,1/1.28,2/.61}{
    \foreach \i/\angle in {0/90,1/210,2/330}{
      \coordinate (v\layer\i) at (\angle:\radius);
    }
    \draw (v\layer0)--(v\layer1)--(v\layer2)--cycle;
  }
  \foreach \i in {0,1,2}{
    \draw (v0\i)--(v1\i)--(v2\i);
  }
  \foreach \layer/\startindex in {0/0,1/1,2/0}{
    \foreach \i in {0,1,2}{
      \ifnum\i=\startindex\relax
        \node[occupied] at (v\layer\i) {\i};
      \else
        \node[vertex] at (v\layer\i) {\i};
      \fi
    }
  }
\end{tikzpicture}
\end{minipage}\hspace{1em}%
\begin{minipage}{.5\linewidth}
\centering
\begin{tabular}{lccc@{\qquad}c}
 & \multicolumn{3}{c}{Current indices} & \\
 & \(T_0\) & \(T_1\) & \(T_2\) & Signs\\
\hline
Before & 0 & 1 & 0 & \(+,-\)\\
After  & 0 & 2 & 0 & \(-,+\)
\end{tabular}
\end{minipage}
\caption{From the outside in, the triangles are \(T_0,T_1,T_2\). Filled
vertices show the starting placement. One slide in \(T_1\) changes its
current index from 1 to 2 and swaps the two signs.}\label{fig:column-slide}
\end{figure}

\begin{lemma}\label{lem:column-path}
Consider a legal reconfiguration sequence on a path of columns
\(T_0,\ldots,T_\ell\), with one token per column and all slides occurring
within columns. Let \(\mu_0\) and \(\mu\) denote the initial and current
numbers of minus signs, respectively. Then
\begin{equation}\label{eq:path-count}
                 \mu=\mu_0+h(T_\ell)-h(T_0).
\end{equation}
If every internal column is a triangle, then two configurations with the
same occupied vertices at the endpoints can be transformed into one
another by internal slides if and only if their sign sequences contain
the same number of minus signs.
\end{lemma}
\begin{proof}
Initially, both endpoint heights are zero, so~\eqref{eq:path-count} holds.
Each successive slide falls into exactly one of three cases, according
to whether the moved token lies in an internal column, the left endpoint,
or the right endpoint. All indices below are read modulo three.

\emph{Internal column.} A legal slide requires the two neighboring tokens
to have the same index, and therefore exchanges adjacent opposite signs,
\((+,-)\leftrightarrow(-,+)\), as illustrated in
Figure~\ref{fig:column-slide}. Thus neither the minus-sign count nor
either endpoint height changes.

\emph{Left endpoint \(T_0\).} For a forward slide \(a\to a+1\), legality
forces the neighboring token to occupy index \(a-1\). The first sign
changes from minus to plus, decreasing \(\mu\) by one while increasing
\(h(T_0)\) by one.

\emph{Right endpoint \(T_\ell\).} For a forward slide \(a\to a+1\), the
neighboring token again occupies index \(a-1\). The last sign changes
from plus to minus, increasing both \(\mu\) and \(h(T_\ell)\) by one.

Backward slides reverse the corresponding changes. Every slide therefore
preserves~\eqref{eq:path-count}, proving the identity by induction.

If the configurations are connected by internal slides, both endpoint
heights remain zero. Equation~\eqref{eq:path-count} therefore gives equal
minus counts.

Conversely, suppose the minus counts agree. Any adjacent pair \((+,-)\)
can be exchanged with \((-,+)\) by one internal slide: the outer tokens
have the same index, and the intervening triangle permits the slide
between the other two indices. Repeating these exchanges moves all
minuses before all pluses in each configuration, yielding identical sign
sequences.

Starting from the common occupied index at the left endpoint, each sign
uniquely determines the next occupied index modulo three. Thus both
configurations reach the same configuration. Reversing the slides from
the second gives the required connection.
\end{proof}

By repeatedly swapping adjacent opposite signs, any sign present can be
brought to either end of the sign sequence while keeping both endpoint
tokens fixed. When the path is part of a larger graph, these slides
additionally require that each destination vertex have no occupied
neighbor outside the path.

\paragraph{Forbidding simultaneous occupancy of three vertices.}

Let \(z_1,z_2,z_3\) be distinct vertices of a graph. Add a triangle on
three new vertices \(b_1,b_2,b_3\). Join \(b_i\) to \(z_i\) for each
\(i\in\{1,2,3\}\). These are the only edges from the triangle to the
original graph. We use the triangle with one token.

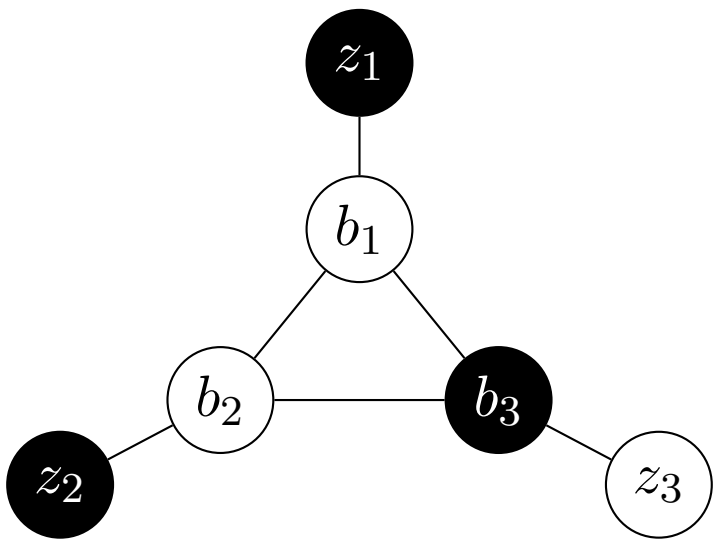
\begin{figure}[!ht]
\centering
\begin{tikzpicture}[
  vertex/.style={circle,draw,fill=white,inner sep=0pt,minimum size=7mm},
  occupied/.style={vertex,fill=black,text=white}]
  \node[vertex]   (b1) at ( 0,     .65) {\(b_1\)};
  \node[vertex]   (b2) at (-.92,  -.48) {\(b_2\)};
  \node[occupied] (b3) at ( .92,  -.48) {\(b_3\)};
  \node[occupied] (z1) at ( 0,    1.75) {\(z_1\)};
  \node[occupied] (z2) at (-1.98,-1.04) {\(z_2\)};
  \node[vertex]   (z3) at ( 1.98,-1.04) {\(z_3\)};
  \draw (b1)--(b2) (b2)--(b3) (b3)--(b1);
  \draw (b1)--(z1) (b2)--(z2) (b3)--(z3);
\end{tikzpicture}
\caption{The six added edges. Filled vertices are occupied. With tokens
on \(z_1\) and \(z_2\), the triangle token must occupy \(b_3\).}
\label{fig:triangle}
\end{figure}

\begin{observation}\label{obs:triangle}
An independent set of the original graph extends to one with a token on
the triangle exactly when some \(z_i\) is unoccupied. If a legal move
changes the occupancy of at most one \(z_i\), and at least one \(z_i\) is
unoccupied both before and after the move, then, from any valid placement
of the triangle token, the move can be performed after at most one slide
within the triangle.
\end{observation}
\begin{proof}
The triangle token can occupy \(b_i\) exactly when \(z_i\) is unoccupied.
Under the stated assumptions, some \(z_i\) remains unoccupied across the
move. Slide the triangle token to \(b_i\), if necessary, then perform the
original move.
\end{proof}

\section{Core gadgets}\label{sec:core-gadgets}

\subsection{Restricting occupancy to an interval}\label{sec:interval}

Let \(T\) be a triangle, \(u\) a vertex outside \(T\), and
\(0\leq a\leq b\leq D\) integers. We attach a path to \(T\) so that
occupying \(u\) requires \(a\leq h(T)\leq b\).

Set \(m=3(D+1)\). Take a path of columns of length \(2m+(b-a)\) ending
at \(T\). All other columns are new. The left endpoint is a single vertex
of index \(-a-b\) modulo three, and every internal column is a triangle.
Each column holds one token. Starting at the left endpoint, the
alternating signs \(+,-,\ldots,+,-\) determine one vertex in each of
the first \(2m+1\) columns. In each of these columns, join \(u\) to
every vertex except that one. Add no other edges from \(u\) to the path.

\begin{figure}[!ht]
\centering
\begin{tikzpicture}[
  font=\small,
  vertex/.style={circle,draw,fill=white,inner sep=0pt,minimum size=5.2mm},
  token/.style={vertex,fill=black,text=white},
  matching/.style={draw=black!62,preaction={draw=white,line width=2.3pt}},
  contact/.style={draw=blue!55!black,densely dashed}]
  \def\columns{1/1.7/0/2/1/2,2/3.8/2/1/0/1,29/7.2/0/2/1/0,%
               30/9.3/2/1/0/1,31/11.4/1/0/2/2,32/13.5/1/2/0/0}
  \foreach \panel/\offset in {0/0,1/-5.2}{
    \begin{scope}[shift={(0,\offset)},name prefix=panel\panel-]
      \coordinate (pin) at (0,-.45);
      \coordinate (u) at (5.5,3.55);
      \foreach \j/\x/\leftindex/\rightindex/\lowerindex/\initialindex in \columns{
        \coordinate (c\j-\leftindex)  at (\x-.52,.7);
        \coordinate (c\j-\rightindex) at (\x+.52,.7);
        \coordinate (c\j-\lowerindex) at (\x,-.45);
        \draw (c\j-\leftindex)--(c\j-\rightindex)--(c\j-\lowerindex)--cycle;
      }
      \draw[matching] (pin)--(c1-0);
      \foreach \j/\k/\common in {1/2/2,29/30/2,30/31/1,31/32/1}{
        \foreach \i in {0,1,2}{
          \ifnum\i=\common\relax
            \ifnum\j=31\relax
              \draw[matching] (c\j-\i) to[bend left=52] (c\k-\i);
            \else
              \draw[matching] (c\j-\i) to[bend left=25] (c\k-\i);
            \fi
          \else
            \draw[matching] (c\j-\i)--(c\k-\i);
          \fi
        }
      }
      \foreach \j/\leftindex/\rightindex in {1/0/2,2/2/1,29/0/2,30/2/1}{
        \draw[contact] (u)--(c\j-\leftindex) (u)--(c\j-\rightindex);
      }
      \ifnum\panel=0\relax
        \node[anchor=west] at (-.3,3.55) {\textbf{(a)} Initial, \(h(T)=0\)};
        \node[vertex] at (u) {\(u\)};
      \else
        \node[anchor=west] at (-.3,3.55) {\textbf{(b)} Later, \(h(T)=3\)};
        \node[token] at (u) {\(u\)};
      \fi
      \node[token] at (pin) {0};
      \node[below=2.8mm] at (pin) {\(T_0\)};
      \foreach \j/\x/\leftindex/\rightindex/\lowerindex/\initialindex in \columns{
        \ifnum\panel=0\relax
          \let\occupiedindex\initialindex
        \else
          \let\occupiedindex\lowerindex
        \fi
        \foreach \i in {0,1,2}{
          \ifnum\i=\occupiedindex\relax
            \node[token] at (c\j-\i) {\i};
          \else
            \node[vertex] at (c\j-\i) {\i};
          \fi
        }
        \ifnum\j=32\relax
          \node[below=2.8mm] at (c\j-\lowerindex) {\(T_{32}=T\)};
        \else
          \node[below=2.8mm] at (c\j-\lowerindex) {\(T_{\j}\)};
        \fi
      }
      \node at (5.5,.12) {\Large\(\cdots\)};
    \end{scope}
  }
\end{tikzpicture}
\caption{The same interval gadget for \([2,4]\) in two configurations.
Filled vertices are occupied. The dots omit \(T_3,\ldots,T_{28}\) and
their incident edges. In (a), \(u\) is empty and the first 13 signs are
minus, followed by 19 plus signs. In (b), \(u\) is occupied. Its dashed
edges force the first 30 signs to alternate. The last two signs are
\(-,+\), read from the indices \(0,2,0\) of \(T_{30},T_{31},T\).
There are now \(15+1=16\) minus signs, so \(h(T)=16-13=3\) by
Lemma~\ref{lem:column-path}.}\label{fig:interval}
\end{figure}
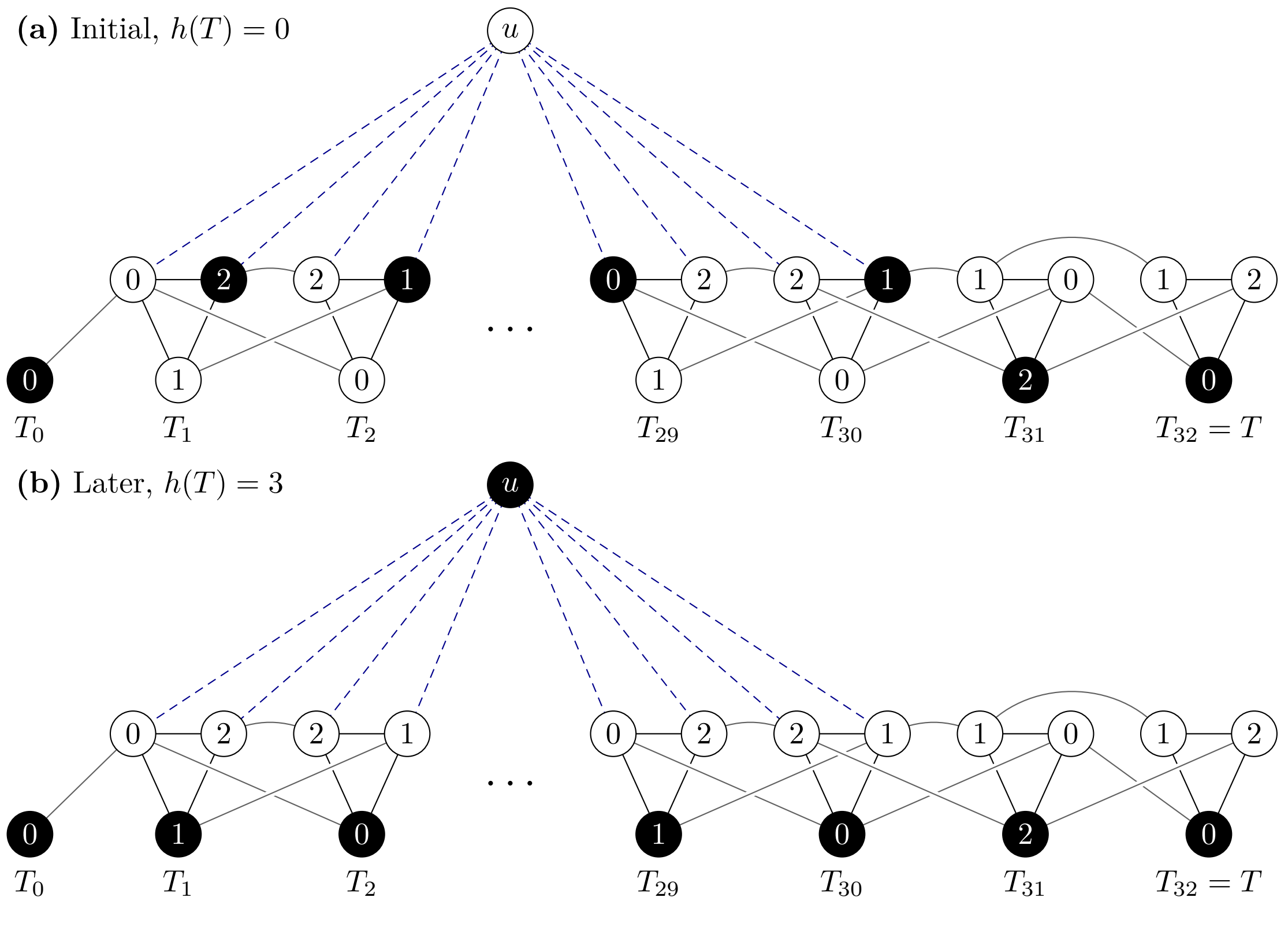

Initially \(u\) is unoccupied. Put all \(m-a\) minus signs first and
fill the remaining positions with plus signs. This gives \(m+b\) plus
signs, so the net change in index along the path is \(a+b\). Thus the
token on \(T\) starts at index zero and \(h(T)=0\).
Figure~\ref{fig:interval} compares this initial configuration with one
at \(h(T)=3\) for the interval \([2,4]\) with \(D=4\). While \(u\) is
empty, the token on \(T\) can make three forward slides and the path
tokens can be arranged as in panel~(b), by Lemma~\ref{lem:interval}
below. The path then allows a token from the rest of the graph to move
onto \(u\).

\begin{lemma}\label{lem:interval}
The interval gadget satisfies the following properties:
\begin{enumerate}
    \item \textbf{Occupancy restriction.}
    If $u$ is occupied, then $a\leq h(T)\leq b$.

    \item \textbf{Availability of $u$.}
    If $a\leq h(T)\leq b$ and $u$ is unoccupied,
    internal slides can free all path neighbors of $u$
    while keeping $T$ fixed.

    \item \textbf{Movement of $T$.}
    If $u$ is unoccupied, the path can be prepared
    to permit any forward or backward slide of $T$
    whose current and resulting heights both lie
    in $[0,D]$.
\end{enumerate}
\end{lemma}

\begin{proof}
Let $\mu$ denote the current number of minus signs on the
path. The path has $2m+(b-a)$ edges and initially contains
$\mu_0=m-a$ minus signs. Since its left endpoint is fixed,
Lemma~\ref{lem:column-path} gives
\[
    \mu=\mu_0+h(T)=m-a+h(T).
\]

\paragraph{Occupancy restriction.}
Suppose that $u$ is occupied. By the construction of the
gadget, the first $2m$ signs must follow the prescribed
alternating pattern. Hence, exactly $m$ of these signs are
minus. The remaining $b-a$ signs must therefore contain
\[
    \mu-m=h(T)-a
\]
minus signs. Since this suffix has length $b-a$, we must have
\[
    0\le h(T)-a\le b-a,
\]
which is equivalent to $a\le h(T)\le b$.

\paragraph{Making $u$ available.}
Suppose $a \le h(T) \le b$ and $u$ is unoccupied.
We want to rearrange the tokens along the path so that
none occupies a vertex adjacent to $u$, without moving
the token on $T$.

By construction, this is achieved when the first $2m$
signs follow the prescribed alternating pattern
$+,-,\ldots,+,-$, which contains exactly $m$ minus signs.
Since the path currently contains
\[
    \mu = m-a+h(T)
\]
minus signs, the remaining $b-a$ signs must contain
\[
    \mu-m=h(T)-a
\]
minuses.

Such an arrangement is possible because
\[
    0 \le h(T)-a \le b-a.
\]
In other words, the remaining $b-a$ positions can
accommodate exactly the required number of minuses.

The resulting arrangement has $m+(h(T)-a)=\mu$ minus
signs, exactly as in the current configuration.
Therefore, by Lemma~\ref{lem:column-path}, we can
reach it using only internal slides, keeping both
endpoints fixed.

In this arrangement, every token in the constrained
columns occupies its unique vertex not adjacent to
$u$. Hence all path neighbors of $u$ are unoccupied,
and $u$ is available.

This also covers $a=b$: the suffix is empty, and
the required arrangement exists exactly when
$h(T)=a$.

\paragraph{Allowing moves of $T$.}
Finally, suppose $u$ is unoccupied and $0\le h(T)\le D$.
The numbers of minus and plus signs are respectively
\[
    \mu=m-a+h(T)
    \qquad\text{and}\qquad
    2m+(b-a)-\mu=m+b-h(T).
\]
Both are strictly positive, since $m=3(D+1)>D$ and
$0\le a\le b\le D$.

By Lemma~\ref{lem:column-path}, internal slides can
rearrange the signs without moving $T$. In particular,
we can place a plus sign at the right end to permit a
forward slide of $T$, or a minus sign there to permit a
backward slide.

Therefore, the path can be prepared for either type of
slide whenever the current and resulting heights of $T$
belong to $[0,D]$.
\end{proof}

The gadget has \(O(D+1)\) vertices and edges.

\subsection{A five-cycle between two triangles}\label{sec:five-cycle}

This gadget connects triangles \(X=\{x_0,x_1,x_2\}\) and
\(Y=\{y_0,y_1,y_2\}\) to a shared vertex \(r\). When \(r\) is occupied,
the difference \(h(Y)-h(X)\) must be either zero or three. We first
prove this restriction and then give slides realizing both values.

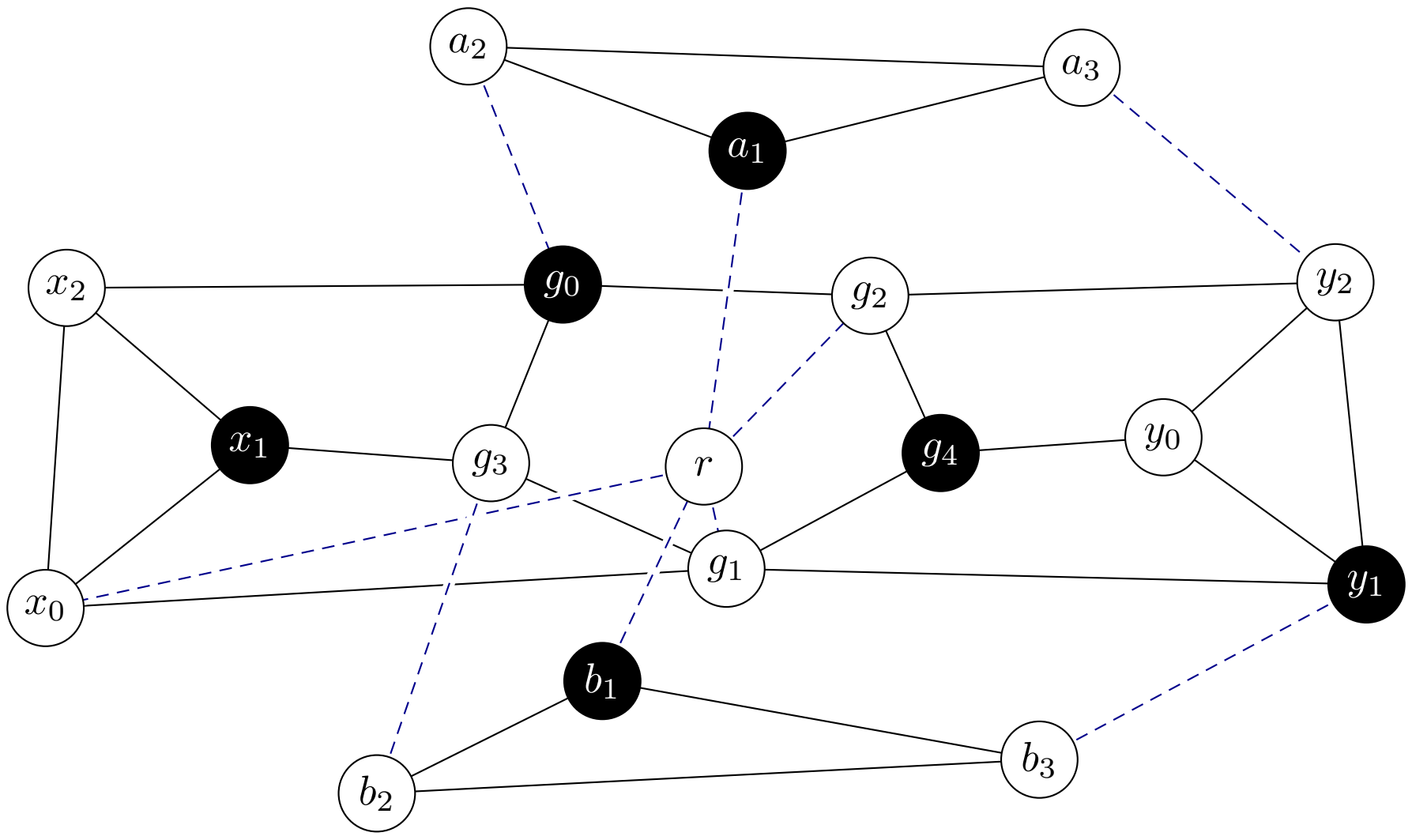
\begin{figure}[!ht]
\centering
\begin{tikzpicture}[
  x=1.18cm,y=.94cm,font=\small,
  vertex/.style={circle,draw,fill=white,inner sep=0pt,minimum size=6.4mm},
  token/.style={vertex,fill=black,text=white},
  contact/.style={draw=blue!55!black,densely dashed,
                  preaction={draw=white,line width=2.6pt}}]
  \foreach \v/\x/\y in {
    x0/-4.65/-1.70,x1/-3.20/-.25,x2/-4.50/1.15,
    g0/-.98/1.18,g2/1.20/1.08,g4/1.70/-.32,g1/.18/-1.35,g3/-1.49/-.41,
    y0/3.28/-.18,y1/4.72/-1.49,y2/4.50/1.20,r/.02/-.44,
    a1/.33/2.37,a2/-1.65/3.30,a3/2.70/3.11,
    b1/-.70/-2.35,b2/-2.30/-3.35,b3/2.40/-3.05}{
      \coordinate (\v) at (\x,\y);
  }
  \draw (x0)--(x1)--(x2)--cycle (y0)--(y1)--(y2)--cycle;
  \draw (a1)--(a2)--(a3)--cycle (b1)--(b2)--(b3)--cycle;
  \draw (g0)--(g2)--(g4)--(g1)--(g3)--cycle;
  \draw (x0)--(g1) (x1)--(g3) (x2)--(g0)
        (y0)--(g4) (y1)--(g1) (y2)--(g2);
  \foreach \u/\v in {r/x0,r/g1,r/g2,a1/r,a2/g0,a3/y2,b1/r,b2/g3,b3/y1}{
    \draw[contact] (\u)--(\v);
  }
  \foreach \v/\name in {x0/x_0,x2/x_2,y0/y_0,y2/y_2,g1/g_1,g2/g_2,g3/g_3,
                         a2/a_2,a3/a_3,b2/b_2,b3/b_3,r/r}{
    \node[vertex] at (\v) {\(\name\)};
  }
  \foreach \v/\name in {x1/x_1,y1/y_1,g4/g_4,g0/g_0,a1/a_1,b1/b_1}{
    \node[token] at (\v) {\(\name\)};
  }
\end{tikzpicture}
\caption{The full five-cycle gadget. Filled vertices show the initial
configuration. Dashed edges are the direct edges from \(r\) and the
attachments of the two auxiliary triangles.}\label{fig:five-cycle}
\end{figure}

Add a five-cycle with vertices in the cyclic order
\(g_0,g_2,g_4,g_1,g_3\). Join \(x_0,x_1,x_2\) to \(g_1,g_3,g_0\),
respectively, and \(y_0,y_1,y_2\) to \(g_4,g_1,g_2\), respectively.
Join \(r\) to \(x_0,g_1,g_2\). Finally, add two copies of the triangle
from Observation~\ref{obs:triangle}. The vertices \(a_1,a_2,a_3\) of
the first copy have outside neighbors \(r,g_0,y_2\), respectively. The
vertices \(b_1,b_2,b_3\) of the second have outside neighbors
\(r,g_3,y_1\), respectively. Figure~\ref{fig:five-cycle} shows all edges.

We consider reconfiguration sequences in which each triangle keeps one
token and the cycle keeps two. A token entering \(r\) therefore comes
from the surrounding graph. The cycle's
independent pairs are \(R_k=\{g_{k-1},g_k\}\), with subscripts modulo
five. Sliding the token on \(g_{k-1}\) to \(g_{k+1}\) changes \(R_k\)
to \(R_{k+1}\). Initially \(r\) is empty and the occupied vertices are
\(x_1,y_1,g_4,g_0,a_1,b_1\).

Let $i,j$ be the current indices of $X,Y$. The table lists all choices
compatible with each cycle pair. Every listed combination is available
with $r$ empty and the auxiliary tokens on $a_1,b_1$. For example,
$R_0=\{g_4,g_0\}$ rules out $x_2$ and $y_0$. To track the
height difference, let $f(i,j)$ be the residue of $j-i$ in
$\{0,1,2\}$, except that $f(2,2)=3$. The last column gives its
values, with rows indexed by $i$ and columns by $j$ in the orders
listed.
\[
\renewcommand{\arraystretch}{1.08}
\begin{array}{c|cc|c}
 \text{Pair}& i & j & f(i,j)\\ \hline
 R_0 & 0,1 & 1,2 & \begin{pmatrix}1&2\\0&1\end{pmatrix}\\[2pt]
 R_1 & 1   & 2,0 & \begin{pmatrix}1&2\end{pmatrix}\\[2pt]
 R_2 & 1,2 & 0   & \begin{pmatrix}2\\1\end{pmatrix}\\[2pt]
 R_3 & 2,0 & 0,1 & \begin{pmatrix}1&2\\0&1\end{pmatrix}\\[2pt]
 R_4 & 2,0 & 1,2 & \begin{pmatrix}2&3\\1&2\end{pmatrix}
\end{array}
\]

\begin{lemma}\label{lem:five-cycle-difference}
The five-cycle gadget satisfies the following properties:
\begin{enumerate}
    \item \textbf{Height-difference invariant.}
    Along every reconfiguration sequence from the initial
    configuration, if the tokens of $X$ and $Y$ occupy
    indices $i$ and $j$, respectively, then
    \[
        h(Y)-h(X)=f(i,j).
    \]

    \item \textbf{Restriction when $r$ is occupied.}
    If $r$ is occupied, the only possible triples
    $(i,j,R_k)$ are
    \[
        (1,1,R_0)
        \qquad\text{and}\qquad
        (2,2,R_4).
    \]
    Consequently,
    \[
        h(Y)-h(X)\in\{0,3\}.
    \]
\end{enumerate}
\end{lemma}

\begin{proof}
We prove the two properties below:

\paragraph{Height-difference invariant.}
We prove the identity by induction on the number of
slides in a legal reconfiguration sequence.

Initially, both triangles have height zero and their
tokens occupy index $1$. Hence,
\[
    h(Y)-h(X)=0=f(1,1).
\]

Now suppose the identity holds before a legal slide.
We show that it continues to hold afterwards.
Every slide either leaves both triangles unchanged,
moves the token in $X$, or moves the token in $Y$.
The compatibility table lists all possible index
combinations for each cycle state, so it suffices
to verify these cases.

If a token moves within the five-cycle or an
auxiliary triangle, or if $r$ changes occupancy,
neither $X$ nor $Y$ moves. Thus their heights
and occupied indices remain unchanged, and
the identity is preserved.

Suppose instead that the token in $Y$ slides
forward. Then $h(Y)-h(X)$ increases by one.
In each row of the compatibility table, the
allowed indices of $Y$, when there are two,
are consecutive in forward order. Moreover,
the corresponding entries of $f$ increase
by one from left to right. Thus every legal
forward slide in $Y$ increases $f(i,j)$ by
exactly one. This also covers wraparound
moves: for example, in state $R_1$, the
slide $y_2\to y_0$ changes
\[
    f(1,2)=1
    \quad\text{to}\quad
    f(1,0)=2.
\]

Similarly, a forward slide in $X$ increases
$h(X)$ by one, so $h(Y)-h(X)$ decreases
by one. In each column of the compatibility
table, the allowed indices of $X$, when
there are two, are consecutive in forward
order, and the corresponding entries of
$f$ decrease by one from top to bottom.
Thus every legal forward slide in $X$
decreases $f(i,j)$ by exactly one.
For example, in state $R_4$ with $j=2$,
the slide $x_2\to x_0$ changes
\[
    f(2,2)=3
    \quad\text{to}\quad
    f(0,2)=2.
\]
This also explains the exceptional value
$f(2,2)=3$.

Backward slides reverse the corresponding
forward moves, so they change both sides
of the identity by the opposite amount
and preserve it as well.

These cases exhaust all legal slides.
In every case, the change in $f(i,j)$
equals the change in $h(Y)-h(X)$.
Therefore, the identity is preserved
after each slide, completing the induction
and proving (1).

\paragraph{Restriction when $r$ is occupied.}
Suppose that $r$ is occupied. Since $r$ is adjacent
to $g_1$, $g_2$, and $x_0$, these three vertices
must be unoccupied.

The five-cycle contains two tokens, so excluding
$g_1$ and $g_2$ leaves only two possible cycle
configurations:
\[
    R_0=\{g_4,g_0\}
    \qquad\text{or}\qquad
    R_4=\{g_3,g_4\}.
\]

First, suppose the cycle is in state $R_0$.
By the compatibility table, the token of $X$ can
occupy only index $0$ or $1$. Since $x_0$ is
adjacent to $r$, it must occupy index $1$.

The token of $Y$ can occupy only index $1$ or $2$.
However, the first auxiliary triangle prevents
$r$, $g_0$, and $y_2$ from being occupied
simultaneously: otherwise, all three vertices of
that auxiliary triangle would be blocked.
Since $r$ and $g_0$ are occupied, $y_2$ must
be unoccupied. Thus $j=1$, giving
\[
    (i,j,R_k)=(1,1,R_0).
\]

Now suppose the cycle is in state $R_4$.
The compatibility table allows $X$ to occupy
index $2$ or $0$. Again, $x_0$ is blocked by
$r$, so $i=2$.

The token of $Y$ can occupy index $1$ or $2$.
The second auxiliary triangle prevents
$r$, $g_3$, and $y_1$ from being occupied
simultaneously. Since $r$ and $g_3$ are
occupied, $y_1$ must be unoccupied, and
therefore $j=2$. Hence
\[
    (i,j,R_k)=(2,2,R_4).
\]

These are the only possible configurations when
$r$ is occupied. Applying (1), we obtain
\[
    h(Y)-h(X)=
    \begin{cases}
        f(1,1)=0, & \text{if the cycle is in }R_0,\\
        f(2,2)=3, & \text{if the cycle is in }R_4.
    \end{cases}
\]
This completes the proof.
\end{proof}

\begin{lemma}\label{lem:five-cycle-slides}
With $r$ unoccupied, the five-cycle gadget admits the
following reconfiguration sequences starting from $(1,1,R_0)$:
\begin{enumerate}
    \item \textbf{Return sequence.}
    A sequence that restores the initial occupied-vertex
    configuration, using three forward slides in each
    of $X$ and $Y$.

    \item \textbf{Transition sequence.}
    A sequence that reaches $(2,2,R_4)$, using one
    forward slide in $X$ and four in $Y$.
\end{enumerate}
At either endpoint, the auxiliary tokens can be
repositioned to make $r$ available, without moving
$X$, $Y$, or the five-cycle.
\end{lemma}

\begin{proof}
We construct both sequences while keeping $r$ unoccupied
and the auxiliary tokens at $a_1,b_1$.

The sequences are described in the table below.
Each arrow within a row represents a forward slide
in $X$ or $Y$, changing exactly one of the occupied
indices $i,j$. Between consecutive rows of a sequence,
the occupied pair of the five-cycle changes from
$R_k$ to $R_{k+1}$, with $i,j$ fixed.

Recall that $R_k=\{g_{k-1},g_k\}$, with indices taken
modulo five. The transition from $R_k$ to $R_{k+1}$
is realized by sliding the token from $g_{k-1}$
to $g_{k+1}$, which is an edge of the five-cycle.
By the compatibility table, every index pair
appearing in a row is allowed in that cycle state.
Moreover, the index pair at each transition between
rows is compatible with both cycle states.
Consequently, every displayed move is legal.

\[
\begin{array}{c|c|c}
  \text{Cycle pair}
  & \text{Return to the start}
  & \text{End at }(2,2,R_4)\\ \hline
  R_0 & \multicolumn{2}{c}{(1,1)\to(1,2)}\\
  R_1 & \multicolumn{2}{c}{(1,2)\to(1,0)}\\
  R_2 & \multicolumn{2}{c}{(1,0)\to(2,0)}\\
  R_3 & (2,0)\to(2,1)\to(0,1)
      & (2,0)\to(2,1)\\
  R_4 & (0,1)
      & (2,1)\to(2,2)\\
  R_0 & (0,1)\to(1,1)
      & \text{---}
\end{array}
\]

In the return sequence, both $X$ and $Y$ make
three forward slides, returning their tokens
to index $1$. Thus, the initial occupied-vertex
configuration is restored, although both heights
have increased by three. In particular, their
height difference remains zero.

In the transition sequence, $X$ makes one forward
slide and $Y$ makes four. The resulting configuration
is $(2,2,R_4)$, with
\[
    h(Y)-h(X)=4-1=3.
\]

It remains to show that $r$ can be made available
at either endpoint.

At $(1,1,R_0)$, slide the auxiliary tokens from
$a_1$ to $a_3$ and from $b_1$ to $b_2$.
These slides are legal because the external
neighbors $y_2$ and $g_3$ of the respective
destinations are unoccupied.

At $(2,2,R_4)$, instead slide the auxiliary
tokens from $a_1$ to $a_2$ and from $b_1$
to $b_3$. These slides are legal because
the external neighbors $g_0$ and $y_1$
of the respective destinations are unoccupied.

In both cases, the auxiliary tokens no longer
block $r$. Furthermore, the vertices
$x_0,g_1,g_2$, which are the remaining
neighbors of $r$ within the gadget, are
unoccupied in both endpoint configurations.
Hence all neighbors of $r$ within the gadget
are unoccupied, and $r$ can be occupied
without moving $X$, $Y$, or the five-cycle.

Once $r$ is vacated, the auxiliary slides
can be reversed.
\end{proof}

\subsection{A chain of five-cycles}\label{sec:chain}

\begin{figure}[!ht]
\centering
\begin{tikzpicture}[
  x=.91cm,y=1.02cm,font=\small,
  vertex/.style={circle,draw,fill=white,inner sep=0pt,minimum size=5.5mm},
  token/.style={vertex,fill=black,text=white},
  contact/.style={preaction={draw=white,line width=2.8pt}},
  bracket/.style={draw=black!60,thin}]
  \coordinate (d) at (-.75,-.97);
  \foreach \k/\c/\s in {0/1/1,1/6.5/1,2/12/1,3/14.6/-1}{
    \foreach \i/\dx/\dy in {0/-.25/-.84,1/.26/0,2/-.21/.89}{
      \coordinate (t\k\i) at ({\c+\s*\dx},\dy);
    }
    \draw (t\k0)--(t\k1)--(t\k2)--cycle;
  }
  \foreach \k/\c/\next in {0/3.7/1,1/9.2/2}{
    \foreach \i/\dx/\dy in {0/-.60/.80,2/.66/.84,4/1.1/-.17,
                              1/.12/-.92,3/-1.05/-.14}{
      \coordinate (g\k\i) at ({\c+\dx},\dy);
    }
    \draw (g\k0)--(g\k2)--(g\k4)--(g\k1)--(g\k3)--cycle;
    \foreach \i/\leftend/\rightend in {0/1/4,1/3/1,2/0/2}{
      \draw[contact] (t\k\i)--(g\k\leftend);
      \draw[contact] (g\k\rightend)--(t\next\i);
    }
    \foreach \i/\sty in {0/token,1/vertex,2/vertex,3/vertex,4/token}{
      \node[\sty] at (g\k\i) {\(g_\i\)};
    }
  }
  \draw (d)--(t00);
  \foreach \i in {0,1,2}{\draw (t2\i)--(t3\i);}
  \node[token] at (d) {\(d\)};
  \foreach \k in {0,1,2}{
    \foreach \i/\sty in {0/vertex,1/token,2/vertex}{
      \node[\sty] at (t\k\i) {\(\i\)};
    }
  }
  \foreach \i/\sty in {0/token,1/vertex,2/vertex}{
    \node[\sty] at (t3\i) {\(\i\)};
  }
  \draw[bracket] (.96,1.33)--(.96,1.48)--(6.42,1.48)--(6.42,1.33);
  \draw[bracket] (6.58,1.33)--(6.58,1.48)--(12.04,1.48)--(12.04,1.33);
  \node[above] at (3.69,1.48) {copy 1};
  \node[above] at (9.31,1.48) {copy 2};
  \foreach \c/\name in {1/T_0,6.5/T_1,12/T_2,14.6/P}{
    \node at (\c,-1.47) {\(\name\)};
  }
\end{tikzpicture}
\caption{Two copies share the triangle \(T_1\), vertex for vertex.
Each \(g_i\) belongs to the copy above it. Numerals on triangles are indices. Filled vertices are initially
occupied. The common vertex \(r\), the auxiliary triangles, and their
incident edges are omitted.}
\label{fig:chain}
\end{figure}
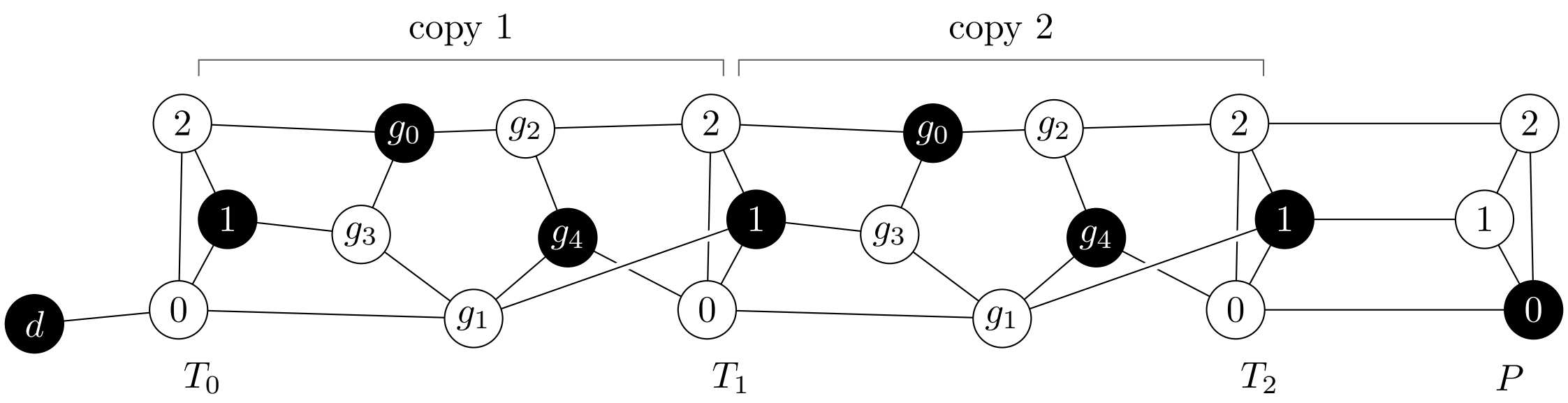

For \(m\geq 1\), take \(m\) copies of the gadget in
Section~\ref{sec:five-cycle}, all with the same vertex \(r\). Write
\(T_0,\ldots,T_m\) for their triangles so that copy \(j\) has
\(X=T_{j-1}\) and \(Y=T_j\). The right triangle of copy \(j\) and the
left triangle of copy \(j+1\) are a single triangle \(T_j\), with the
same vertex at each index. The five-cycle and both auxiliary triangles are private
to each copy. Figure~\ref{fig:chain} shows how two copies fit together.

At the left end, add a singleton column \(d\) adjacent only to the
vertex of index zero in \(T_0\). At the right end, add a triangle
\(P=\{p_0,p_1,p_2\}\) and join vertices of the same index in \(T_m\)
and \(P\). Join \(r\) to \(p_1\) and \(p_2\). Initially each copy is
in the configuration of Figure~\ref{fig:five-cycle}, \(d\) and \(p_0\)
are occupied, and \(r\) is empty. Throughout, \(d\) stays occupied,
each triangle holds one token, and each five-cycle holds two.
The chain has \(O(m)\) vertices and edges, since each copy has constant size.

\begin{lemma}\label{lem:chain-heights}
Whenever $r$ is occupied, $h(P)$ is either $0$ or $3m$.
\end{lemma}

\begin{proof}
Suppose $r$ is occupied. By
Lemma~\ref{lem:five-cycle-difference}, each five-cycle
copy must be in one of two configurations:
$(1,1,R_0)$ or $(2,2,R_4)$.

Since consecutive copies share a triangle, they must
all be in the same configuration. Indeed, if one copy
is in $(1,1,R_0)$, its right triangle occupies index
one, forcing the next copy into the same configuration.
Similarly, if one copy is in $(2,2,R_4)$, its right
triangle occupies index two, forcing the next copy
into that configuration.

We therefore have two cases.

\medskip
\noindent\textbf{Case 1: All copies are in $(1,1,R_0)$.}
By Lemma~\ref{lem:five-cycle-difference}, each copy
satisfies
\[
    h(T_j)-h(T_{j-1})=0.
\]
Summing over $j=1,\ldots,m$ gives
\[
    h(T_m)=h(T_0).
\]
The occupied singleton $d$ prevents the token of
$T_0$ from occupying index zero. Since it starts at
index one, its height is zero whenever it occupies
index one. Thus $h(T_0)=h(T_m)=0$.

\medskip
\noindent\textbf{Case 2: All copies are in $(2,2,R_4)$.}
Each copy now satisfies
\[
    h(T_j)-h(T_{j-1})=3.
\]
Summing gives
\[
    h(T_m)=h(T_0)+3m.
\]
Again, the occupied singleton $d$ blocks index zero
in $T_0$. Since its token starts at index one, its
height is one whenever it occupies index two.
Therefore,
\[
    h(T_m)=3m+1.
\]

\medskip
It remains to determine $h(P)$. Since $r$ is
adjacent to $p_1$ and $p_2$, the token on $P$
must occupy $p_0$.

The edge between $T_m$ and $P$ initially has
a minus sign, corresponding to indices $(1,0)$.
By the path invariant~\eqref{eq:path-count},
\[
    \mu=1+h(P)-h(T_m),
\]
where $\mu\in\{0,1\}$ counts the minus signs
on this single edge.

In Case 1, $T_m$ occupies index one, so the
edge has a minus sign and $\mu=1$. Since
$h(T_m)=0$, we obtain $h(P)=0$.

In Case 2, $T_m$ occupies index two, so the
edge has a plus sign and $\mu=0$. Since
$h(T_m)=3m+1$, we obtain
\[
    h(P)=h(T_m)-1=3m.
\]

Hence, whenever $r$ is occupied,
\[
    h(P)\in\{0,3m\}.
\]
This completes the proof. 
\end{proof}

\begin{lemma}\label{lem:chain-slides}
With $r$ empty, there is a reconfiguration sequence
from the initial configuration in which the token
on $P$ makes exactly $3m$ slides, all forward.
At either end, the auxiliary tokens can move to
free all neighbors of $r$ in the chain while
$P$ stays fixed.
\end{lemma}

\begin{proof}
We first construct a reconfiguration sequence
for the chain of $m$ five-cycle copies, without
yet considering the moves of $P$.

We prove by induction on $j$, for
$1\le j\le m$, that the first $j$ copies admit
a reconfiguration sequence satisfying:
\begin{enumerate}
    \item The vertex $r$ remains unoccupied,
    and every copy ends in configuration
    $(2,2,R_4)$.

    \item For every $0\le k\le j$, the token
    on $T_k$ makes exactly $3k+1$ forward
    slides and no backward slides.
\end{enumerate}

Throughout these sequences, the auxiliary tokens
remain in their initial positions.

\medskip
\noindent\textbf{Base case ($j=1$).}
By Lemma~\ref{lem:five-cycle-slides}, the
transition sequence of the first copy starts
at $(1,1,R_0)$ and ends at $(2,2,R_4)$,
using one forward slide in $T_0$ and four
forward slides in $T_1$.

Thus $T_0$ makes $3(0)+1=1$ forward slide,
and $T_1$ makes $3(1)+1=4$ forward slides.
Both properties hold.

\medskip
\noindent\textbf{Inductive step.}
Suppose we have constructed a sequence for the
first $j$ copies satisfying both properties,
where $j<m$.

We now add copy $j+1$, which shares the
triangle $T_j$ with the existing chain.

For this new copy, consider the sequence
obtained by repeating the return sequence of
Lemma~\ref{lem:five-cycle-slides} exactly
$j$ times, followed by its transition sequence.

Each return sequence makes three forward slides
in both terminal triangles, while the transition
sequence makes one forward slide in the left
triangle and four in the right triangle.
Therefore, in the new copy, the total numbers
of forward slides are
\[
\begin{aligned}
    T_j &: 3j+1,\\
    T_{j+1} &: 3j+4=3(j+1)+1.
\end{aligned}
\]

By the inductive hypothesis, the sequence for
the first $j$ copies also makes exactly
$3j+1$ forward slides in $T_j$.

Hence both sequences require the same number
of slides on their shared triangle $T_j$.
Moreover, both start with its token at index
one, and every such slide is forward.
Thus, their successive slides on $T_j$
are identical.

We synchronize the two sequences as follows.
Execute the private moves of each sequence
up to its next slide in $T_j$. Then perform
that shared slide exactly once, and continue
with the private moves of both sequences
until their next shared slide. Repeat this
process, and finally execute any remaining
private moves.

This interleaving preserves the order of all
moves in each original sequence. Between
successive shared slides, $T_j$ remains fixed,
$r$ is unoccupied, and no edge joins private
vertices belonging to different copies.
Therefore, every private move remains legal.
Each shared slide is also legal because both
sequences have reached the configurations
required for that slide.

Consequently, the combined sequence is legal.
The first $j$ copies retain their required
final configurations and move counts, while
copy $j+1$ reaches $(2,2,R_4)$ and its new
terminal triangle $T_{j+1}$ makes exactly
$3(j+1)+1$ forward slides.

This completes the inductive step.

\medskip
By induction, the full chain admits a sequence
in which every copy reaches $(2,2,R_4)$
and, for each $0\le k\le m$, the token
on $T_k$ makes exactly $3k+1$ forward slides.
In particular, $T_m$ makes $3m+1$ forward slides.
The only slide of $T_0$ is from index one
to index two, as required by the occupied
singleton $d$.

\medskip
\noindent\textbf{Extending the sequence to $P$.}
We now interleave forward slides of $P$ with
the constructed sequence for the chain.

Initially, the tokens on $T_m$ and $P$
occupy indices $(1,0)$. Let $T_m$ make
its first forward slide, reaching $(2,0)$.

Before each subsequent forward slide of $T_m$,
advance the token on $P$ once. For example,
the first such pair of slides is
\[
    (2,0)\longrightarrow(2,1)
    \longrightarrow(0,1).
\]
The same pattern repeats with both indices
increased by one modulo three. Thus, the two
tokens always occupy distinct indices,
and the matching edges between $T_m$ and
$P$ never prevent these slides.

Since $T_m$ makes $3m+1$ forward slides,
the token on $P$ makes exactly $3m$
forward slides, all forward. As $3m$ is
divisible by three, $P$ ends at index zero,
that is, at $p_0$.

\medskip
\noindent\textbf{Making $r$ available.}
At the beginning of the sequence, all copies
are in configuration $(1,1,R_0)$, while
at the end they are all in configuration
$(2,2,R_4)$. The token on $P$ occupies
$p_0$ at both endpoints.

By Lemma~\ref{lem:five-cycle-slides},
the auxiliary tokens of each copy can be
repositioned to free all its neighbors of
$r$, without moving its terminal triangles
or five-cycle.

Since these auxiliary triangles are private
to their respective copies, the preparations
can be performed independently. Furthermore,
$P$ occupies $p_0$, so neither $p_1$ nor
$p_2$ blocks $r$.

Hence, at either endpoint, all neighbors of
$r$ within the chain can be made unoccupied
while $P$ remains fixed. Once $r$ is vacated,
the auxiliary slides can be reversed.
\end{proof}

\subsection{Excluding one chosen height}\label{sec:exclude}

Fix integers \(B,t\) with \(0<t<B\) and \(t\) divisible by three.
Let \(T\) be a triangle initially at index zero. We attach \(T\) to
the chain so that when \(h(T)=t\), \(h(P)\) lies strictly between the
two heights allowed by the chain when \(r\) is occupied.

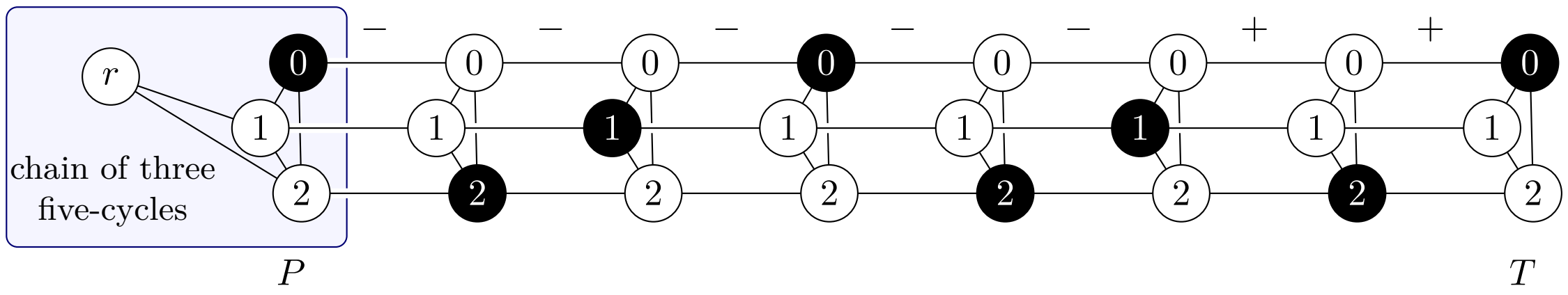
\begin{figure}[!ht]
\centering
\begin{tikzpicture}[
  x=1cm,y=1cm,font=\small,
  vertex/.style={circle,draw,fill=white,inner sep=0pt,minimum size=5.4mm},
  token/.style={vertex,fill=black,text=white}]
  \filldraw[fill=blue!4,draw=blue!45!black,rounded corners=3pt]
    (-2.66,-1.13) rectangle (.57,1.15);
  \node[align=center,font=\footnotesize] at (-1.65,-.60)
    {chain of three\\five-cycles};
  \coordinate (r) at (-1.67,.49);
  \foreach \j in {0,...,7}{
    \foreach \i/\dx/\dy in {0/.11/.62,1/-.25/0,2/.14/-.62}{
      \coordinate (v\j\i) at ({1.67*\j+\dx},\dy);
    }
    \draw (v\j0)--(v\j1)--(v\j2)--cycle;
  }
  \draw (r)--(v01) (r)--(v02);
  \foreach \j in {1,...,7}{
    \pgfmathtruncatemacro{\prev}{\j-1}
    \foreach \i in {0,1,2}{
      \draw[preaction={draw=white,line width=2.7pt}] (v\prev\i)--(v\j\i);
    }
  }
  \node[vertex] at (r) {\(r\)};
  \foreach \j/\occ in {0/0,1/2,2/1,3/0,4/2,5/1,6/2,7/0}{
    \foreach \i in {0,1,2}{
      \ifnum\i=\occ\relax
        \node[token] at (v\j\i) {\(\i\)};
      \else
        \node[vertex] at (v\j\i) {\(\i\)};
      \fi
    }
  }
  \foreach \j/\sign in {1/-,2/-,3/-,4/-,5/-,6/+,7/+}{
    \node at ({1.67*(\j-.5)},.94) {\(\sign\)};
  }
  \node at (.04,-1.37) {\(P\)};
  \node at (11.73,-1.37) {\(T\)};
\end{tikzpicture}
\caption{The added path for \(B=4\) and \(t=3\). The box shows only
\(r\) and \(P\) from the chain. Filled vertices are initially occupied.
The five minus signs and two plus signs are read from \(P\) to \(T\).}
\label{fig:exclude}
\end{figure}

Let \(L\) be the least multiple of three at least \(B+3\). Take the
chain of \(L/3\) five-cycles from Section~\ref{sec:chain}. From \(P\)
to \(T\) add a path of \(S=L-2\) edges, with a triangle at each column
and one token per triangle. The chain and the path share only \(P\).
In particular, \(r\) has no new neighbors. Initially give the path
\(\mu_0=L-1-t\) minus signs followed by plus signs. This is possible
since \(0\leq\mu_0\leq S\). Starting at index zero in \(P\), the
signs lead to index zero in \(T\) because their sum is
\(S-2\mu_0=2t-L\), a multiple of three. The construction has \(O(B)\)
vertices and edges. Figure~\ref{fig:exclude} shows the path for \(t=3\).

\begin{lemma}\label{lem:exclude-height}
If $h(T)=t$, then $r$ is unoccupied.
\end{lemma}
\begin{proof}
Let $\mu$ denote the number of minus signs on the added
path, which has $L-2$ edges. Since the path initially
contains $\mu_0=L-1-t$ minus signs, the path invariant
\eqref{eq:path-count} gives
\[
    h(P)+\mu=\mu_0+h(T)=L-1-t+h(T).
\]

Suppose $h(T)=t$. Then
\[
    h(P)+\mu=L-1.
\]
Since the path has $L-2$ edges, we have
$0\leq\mu\leq L-2$, and therefore
\[
    1\leq h(P)=L-1-\mu\leq L-1.
\]

However, Lemma~\ref{lem:chain-heights} states that
whenever $r$ is occupied, $h(P)\in\{0,L\}$.
Neither value lies in $[1,L-1]$.
Hence $r$ cannot be occupied when $h(T)=t$.
\end{proof}

\begin{lemma}\label{lem:exclude-slides}
With $r$ unoccupied, the excluded-height gadget admits
a reconfiguration sequence in which $T$ advances from
height $0$ to $B$ using only forward slides.
Moreover, at every height other than $t$, the sequence
reaches a configuration in which slides fixing $T$
can free all neighbors of $r$ within the gadget.
\end{lemma}

\begin{proof}
Let $\mu$ denote the number of minus signs on the
path from $P$ to $T$, which has $S=L-2$ edges.
Recall that initially $\mu_0=L-1-t$.

By Lemma~\ref{lem:column-path}, internal slides
can rearrange the signs while keeping both
endpoints fixed. A forward slide in $T$ requires
a plus at the right end and converts it into a
minus, increasing $\mu$ by one. Conversely, a
forward slide in $P$ requires a minus at the left
end and converts it into a plus, decreasing
$\mu$ by one.

The idea is to keep the chain in its initial
configuration while $T$ advances through heights
$0,\ldots,t-1$, and to complete the chain's
transition to its final configuration by the time
$T$ reaches height $t+1$. By Lemma~\ref{lem:chain-heights},
these endpoint configurations are the only ones
in which $r$ can be occupied. Lemma~8 guarantees
that the neighbors of $r$ within the chain can
be freed at both endpoints. We therefore
coordinate the movements of $P$ and $T$ so
that the transition between these endpoints
occurs while $T$ crosses the excluded height $t$.

We construct the sequence in four phases, keeping
$r$ unoccupied throughout.

\medskip
\noindent\textbf{Phase 1: Advance $T$ to height $t-1$.}
Keep the five-cycle chain fixed, so $h(P)=0$,
and advance $T$ from height $0$ to $t-1$.
Before each forward slide, bring an available
plus sign to the right end.

After these $t-1$ slides, the number of
minus signs is
\[
    \mu=\mu_0+(t-1)
       =L-1-t+t-1
       =L-2=S.
\]
Thus all signs are now minus, and $T$ cannot
advance further while $P$ remains fixed.

\medskip
\noindent\textbf{Phase 2: Advance $P$ to height $S$.}
Keep $T$ fixed at height $t-1$ and begin the
chain reconfiguration sequence provided by
Lemma~\ref{lem:chain-slides}.

Before each of the first $S$ forward slides
in $P$, use internal slides to bring a minus
sign to the left end of the path.

Each slide in $P$ decreases $\mu$ by one.
After $S$ such slides, we obtain
\[
    h(T)=t-1,\qquad h(P)=S,\qquad \mu=0.
\]
All signs are now plus. The chain sequence
still requires two additional forward slides
in $P$, but these cannot be performed until
new minus signs become available.

\medskip
\noindent\textbf{Phase 3: Cross the forbidden height.}
We coordinate the next two forward slides
of $T$ with the remaining two forward slides
of $P$.

Starting from $(h(T),h(P),\mu)=(t-1,S,0)$,
perform the following operations:
\[
\begin{array}{c|ccc}
\text{Operation} & h(T) & h(P) & \mu\\ \hline
\text{Initial}    & t-1 & S   & 0\\
\text{Advance }T  & t   & S   & 1\\
\text{Advance }P  & t   & S+1 & 0\\
\text{Advance }T  & t+1 & S+1 & 1\\
\text{Advance }P  & t+1 & L   & 0
\end{array}
\]

Each forward slide in $T$ creates a minus
sign. Before the subsequent slide in $P$,
we bring this minus to the left end, allowing
$P$ to advance and consume it.

All four endpoint slides are therefore legal.
Moreover, all slides of $T$ are forward.

At the end of this phase,
\[
    h(T)=t+1,\qquad h(P)=L,\qquad \mu=0.
\]

\medskip
\noindent\textbf{Phase 4: Advance $T$ to height $B$.}
Finish any remaining private moves of the
chain sequence while keeping $T$ and $P$
fixed. The chain is now in its final
configuration.

Keep the chain fixed and advance $T$ from
height $t+1$ to $B$. Since all $S$ signs
are initially plus and
\[
    B-t-1\le S,
\]
there are enough plus signs for the remaining
forward slides.

\medskip
\noindent\textbf{Availability of $r$.}
It remains to establish the claimed
checking opportunities.

At every height $0\le h(T)\le t-1$,
Phase 1 reaches a configuration where the
chain is in its initial endpoint state,
with $h(P)=0$.

Similarly, at every height
$t+1\le h(T)\le B$, Phase 4 reaches a
configuration where the chain is in its
final endpoint state, with $h(P)=L$.

By Lemma~\ref{lem:chain-slides}, at either
endpoint state the auxiliary tokens can be
repositioned to free all neighbors of $r$
within the chain, without moving $P$ or $T$.
The added path introduces no additional
neighbors of $r$.

After $r$ is vacated, these auxiliary slides
can be reversed, allowing the constructed
sequence to continue.

Thus the sequence advances $T$ from height
$0$ to $B$ using only forward slides, while
providing the required opportunity to occupy
$r$ at every height other than $t$.
\end{proof}

\section{The variable gadget}\label{sec:variables}

Let \(D\) be a positive multiple of three. All variables share a
triangle \(Q=\{q_0,q_1,q_2\}\), initially at index zero, and a vertex
\(r\), initially empty and adjacent to \(q_1,q_2\). Each variable \(x\)
has its own triangle \(V_x=\{v_0,v_1,v_2\}\), also initially at index
zero. Each triangle and singleton column holds one token throughout.
On the positive part of the passage to \(D\), we check the variable
when \(r\) is occupied. Equal heights in \(V_x\) and \(Q\) represent
false. One extra forward turn in \(V_x\) represents true.

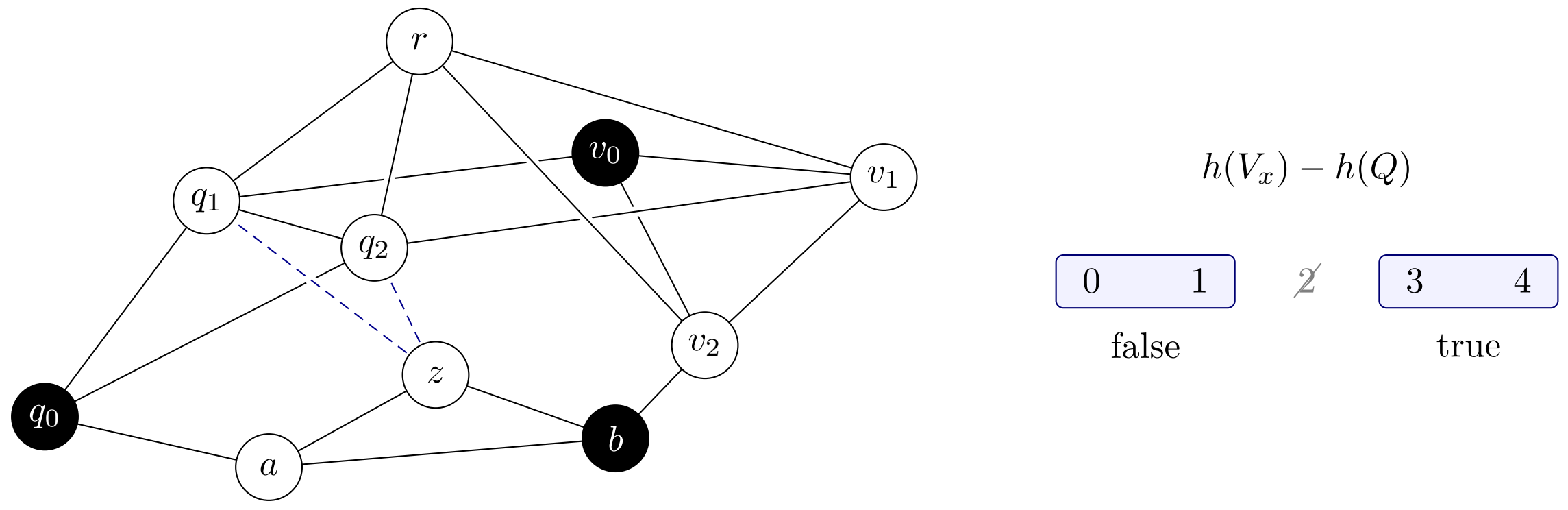
\begin{figure}[!ht]
\centering
\begin{minipage}[c]{.59\linewidth}
\centering
\begin{tikzpicture}[
  x=.97cm,y=.84cm,font=\small,
  vertex/.style={circle,draw,fill=white,inner sep=0pt,minimum size=6.4mm},
  token/.style={vertex,fill=black,text=white},
  contact/.style={preaction={draw=white,line width=2.8pt}}]
  \foreach \v/\x/\y in {q0/-4.60/-1.47,q1/-2.99/1.01,q2/-1.32/.47,
    v0/.98/1.56,v1/3.75/1.28,v2/1.97/-.65,r/-.87/2.84,
    a/-2.37/-2.05,b/1.08/-1.72,z/-.71/-.99}{
    \coordinate (\v) at (\x,\y);
  }
  \draw (q0)--(q1)--(q2)--cycle (v0)--(v1)--(v2)--cycle
        (a)--(b)--(z)--cycle;
  \foreach \u/\v in {q0/a,b/v2,q1/v0,q2/v1,r/q1,r/q2,r/v1,r/v2}{
    \draw[contact] (\u)--(\v);
  }
  \foreach \v in {q1,q2}{
    \draw[contact,densely dashed,blue!55!black] (z)--(\v);
  }
  \foreach \v/\name in {q1/q_1,q2/q_2,v1/v_1,v2/v_2,a/a,z/z,r/r}{
    \node[vertex] at (\v) {\(\name\)};
  }
  \foreach \v/\name in {q0/q_0,v0/v_0,b/b}{
    \node[token] at (\v) {\(\name\)};
  }
\end{tikzpicture}
\end{minipage}\hfill
\begin{minipage}[c]{.38\linewidth}
\centering
\begin{tikzpicture}[x=1.04cm,y=1cm,font=\small]
  \node at (2,1.08) {\(h(V_x)-h(Q)\)};
  \draw[rounded corners=2pt,fill=blue!5,draw=blue!45!black]
    (-.33,-.26) rectangle (1.33,.25)
    (2.67,-.26) rectangle (4.33,.25);
  \foreach \i in {0,1,3,4}{\node at (\i,0) {\(\i\)};}
  \node[gray] at (2,0) {\(2\)};
  \draw[gray] (1.88,-.16)--(2.13,.17);
  \node at (.5,-.62) {false};
  \node at (3.5,-.62) {true};
\end{tikzpicture}
\end{minipage}
\caption{The variable gadget and its two possible pairs of height
differences on the positive part of the passage to \(D\). Filled vertices are initially
occupied. The paths are omitted. The dashed edges are the two edges
from \(z\) to \(Q\) in the interval gadget.}\label{fig:variable}
\end{figure}

Join \(q_1\) to \(v_0\), \(q_2\) to \(v_1\), and \(r\) to \(v_1,v_2\).
Add a triangle \(\{a,b,z\}\), initially occupied at \(b\), and edges
\(aq_0\) and \(bv_2\). Attach the interval gadget of
Section~\ref{sec:interval} to \(Q\) and \(z\) with interval \([0,0]\).
Thus \(z\) can be occupied only when \(h(Q)=0\).
Figure~\ref{fig:variable} shows the edges among these vertices.

Finally, add a path of columns of length \(D+3\) from a singleton
of index zero to \(V_x\), with triangles at all internal columns.
Give the path all plus signs initially.
Since \(D+3\) is divisible by three, the path ends at \(v_0\).
By~\eqref{eq:path-count}, \(h(V_x)\) counts its minus signs, so
\[
                      0\leq h(V_x)\leq D+3.
\]
Only \(Q\) and \(r\) are shared between variables. Each variable adds
\(O(D)\) vertices and edges.

\begin{lemma}\label{lem:variable-residue}
If $h(Q)>0$, then
$
    h(V_x)-h(Q)\not\equiv 2\pmod 3.
$
\end{lemma}

\begin{proof}
Let $i$ and $j$ denote the occupied indices of
$Q$ and $V_x$, respectively. Since both triangles
initially occupy index zero, we have
\[
    h(V_x)-h(Q)\equiv j-i\pmod 3.
\]
Thus, a difference of two modulo three is possible
only for the index pairs
\[
    (i,j)\in\{(0,2),(1,0),(2,1)\}.
\]

The edges $q_1v_0$ and $q_2v_1$ directly forbid
the pairs $(1,0)$ and $(2,1)$, respectively.

For the remaining pair $(0,2)$, the occupied
vertices $q_0$ and $v_2$ block $a$ and $b$
in the auxiliary triangle $\{a,b,z\}$.
Since this triangle contains a token, that
token would be forced onto $z$.

However, the interval gadget $[0,0]$ attached
to $z$ permits $z$ to be occupied only when
$h(Q)=0$, by Lemma~\ref{lem:interval}.
This contradicts $h(Q)>0$.

Therefore, all three index pairs are impossible
at positive height, and
\[
    h(V_x)-h(Q)\not\equiv 2\pmod 3.
\]
This completes the proof. 
\end{proof}

Consider a reconfiguration sequence in which \(h(Q)\) reaches \(D\).
End the passage at the first visit to \(D\) and start it at the last
preceding visit to zero.

\begin{lemma}\label{lem:variable-fixed}
Throughout the positive part of this passage,
$h(V_x)-h(Q)$ remains in one fixed pair, either
$\{0,1\}$ or $\{3,4\}$.
\end{lemma}

\begin{proof}
Let $\Delta=h(V_x)-h(Q)$.

By Lemma~\ref{lem:variable-residue}, whenever
$h(Q)>0$, we have $\Delta\not\equiv 2\pmod 3$.
Thus, the allowed values of $\Delta$ form
disjoint pairs
\[
    \{3j,3j+1\},\qquad j\in \mathbb{Z},
\]
with consecutive pairs separated by a
forbidden integer.

Each token slide changes at most one of
$h(V_x)$ and $h(Q)$, by at most one.
Consequently, $\Delta$ can change by at
most one in a single slide.

Moving between two distinct pairs would
therefore require passing through a value
congruent to $2$ modulo three, which is
forbidden by Lemma~\ref{lem:variable-residue}.
Since $h(Q)$ remains positive throughout
the considered portion of the passage,
$\Delta$ must remain in one fixed pair,
even if backward slides occur.

It remains to determine which pair is possible.

At the first positive height, $h(Q)=1$.
Since the height-bounding path guarantees
$h(V_x)\geq 0$, we have
\[
    \Delta=h(V_x)-h(Q)\geq -1.
\]
Hence the fixed pair cannot have $j<0$,
since such pairs contain only values
at most $-2$.

At the end of the passage, $h(Q)=D$.
The same path guarantees
$h(V_x)\leq D+3$, giving
\[
    \Delta=h(V_x)-h(Q)\leq 3.
\]
Thus the fixed pair cannot have $j\geq 2$,
since such pairs contain only values
at least $6$.

Since the pair remains unchanged throughout
the positive passage, these two bounds imply
$j\in\{0,1\}$.

Therefore, $\Delta$ remains in one fixed
pair, either $\{0,1\}$ or $\{3,4\}$.
\end{proof}

\begin{lemma}\label{lem:variable-choice}
Either truth value can be chosen at $h(Q)=0$ and maintained
while $Q$ advances to height $D$ with $r$ unoccupied.
\end{lemma}
\begin{proof}
We first show how to initialize either truth value.
For false, leave $V_x$ at height zero, so that
$h(V_x)-h(Q)=0$.

For true, keep $Q$ fixed at $q_0$ and use
Lemma~\ref{lem:interval} to free the path neighbors
of $z$. Since $h(Q)=0$, the interval gadget $[0,0]$
allows the auxiliary token to slide from $b$ to $z$.
With $b$ unoccupied, the token in $V_x$ can make
one complete forward turn,
\[
    v_0\to v_1\to v_2\to v_0,
\]
while $Q$ remains at $q_0$ and the auxiliary
token remains at $z$. These slides respect the
cross-edges between $Q$ and $V_x$ and increase
$h(V_x)$ by three. Since $v_2$ is again
unoccupied, the auxiliary token can return
from $z$ to $b$.

Thus, at $h(Q)=0$, we can establish either
\[
    h(V_x)-h(Q)=0
    \qquad\text{or}\qquad
    h(V_x)-h(Q)=3.
\]

We next show how to maintain the chosen difference
while $Q$ advances from height zero to $D$.
The strategy is to advance $V_x$ once before
each forward slide in $Q$. The following table
describes the occupied indices of $Q,V_x$ and
the position of the auxiliary token during
each pair of slides:
\[
\begin{array}{c|cc}
 \text{Index pairs in order}
 & \text{During the slide in }V_x
 & \text{During the slide in }Q\\ \hline
 (0,0)\to(0,1)\to(1,1) & b & b\\
 (1,1)\to(1,2)\to(2,2) & a & a\\
 (2,2)\to(2,0)\to(0,0) & a & b
\end{array}
\]

The auxiliary token switches from $b$ to $a$
at $(1,1)$, and from $a$ to $b$ at $(2,0)$.
Both switches are legal because $q_0$ and
$v_2$ are unoccupied at these configurations.

Every pair of indices appearing in the table
respects the edges $q_1v_0$ and $q_2v_1$.
Furthermore, the auxiliary token never occupies
$z$ during this progression. Thus, by
Lemma~\ref{lem:interval}, the interval path
attached to $z$ can be prepared for every
forward slide in $Q$.

Before each forward slide in $V_x$, its
height-bounding path also permits the required
move. Indeed, this path initially has $D+3$
plus signs, and its minus count equals
$h(V_x)$. Since $h(V_x)<D+3$ before every
requested forward slide, at least one plus
sign remains. By Lemma~\ref{lem:column-path},
internal slides can bring this plus to the
right end. This also applies to the three
initial slides used to choose true.

Let $\delta\in\{0,3\}$ denote the chosen
height difference. Each round first advances
$V_x$, temporarily increasing the difference
from $\delta$ to $\delta+1$, and then advances
$Q$, restoring it to $\delta$. Hence
\[
    \delta\to\delta+1\to\delta.
\]
The chosen difference is therefore maintained
after every forward slide in $Q$.

Finally, consider several variable gadgets
sharing the same triangle $Q$. Before each
forward slide in $Q$, perform the necessary
auxiliary preparations and forward slide in
every private variable triangle $V_x$.
Then prepare the interval paths attached to
$Q$ and perform the shared forward slide
in $Q$ exactly once.

Since distinct variable gadgets have no edges
between their private vertices, these
preparations can be performed independently.
Each variable therefore maintains its chosen
height difference while the shared triangle
$Q$ advances to height $D$, with $r$
remaining unoccupied throughout.
\end{proof}

\section{The clause gadget}\label{sec:clauses}

A clause with interval \([\ell_1,\ell_2]\), where
\(1\leq\ell_1\leq\ell_2\leq D-1\), has its own five-vertex graph with
two tokens. Its initial pair \(\{a,c\}\) records that the clause has
yet to be checked. To reach \(\{d,b\}\), a token must pass through
\(u\). We use the interval gadget to restrict when this can happen.

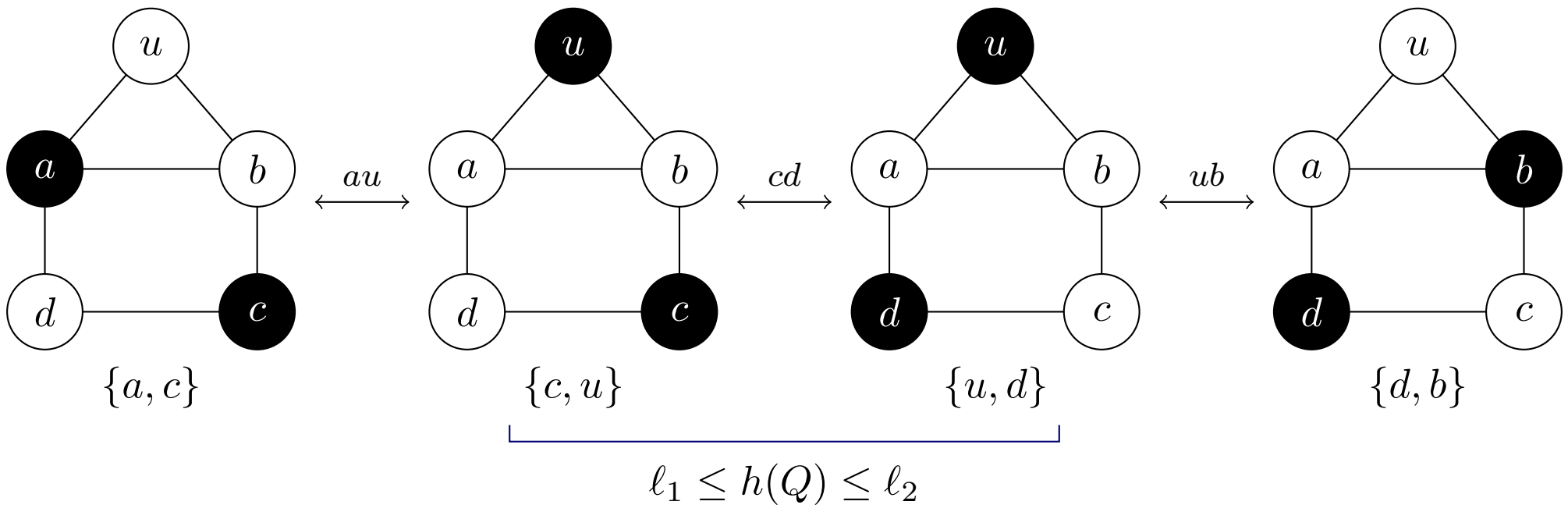
\begin{figure}[!ht]
\centering
\begin{tikzpicture}[
  x=1cm,y=1cm,font=\small,
  vertex/.style={circle,draw,fill=white,inner sep=0pt,minimum size=6.4mm},
  token/.style={vertex,fill=black,text=white}]
  \foreach \j/\astyle/\cstyle/\ustyle/\dstyle/\bstyle in {
    0/token/token/vertex/vertex/vertex,
    1/vertex/token/token/vertex/vertex,
    2/vertex/vertex/token/token/vertex,
    3/vertex/vertex/vertex/token/token}{
    \begin{scope}[shift={({3.58*\j},0)}]
      \coordinate (a) at (-.90,.33);
      \coordinate (b) at (.90,.33);
      \coordinate (c) at (.90,-.88);
      \coordinate (d) at (-.90,-.88);
      \coordinate (u) at (0,1.37);
      \draw (a)--(u)--(b)--(c)--(d)--cycle (a)--(b);
      \foreach \v/\sty in {a/\astyle,c/\cstyle,u/\ustyle,d/\dstyle,b/\bstyle}{
        \node[\sty] at (\v) {\(\v\)};
      }
    \end{scope}
  }
  \foreach \j/\edge in {0/au,1/cd,2/ub}{
    \draw[<->] ({3.58*\j+1.39},.05)--node[above,font=\scriptsize]{\(\edge\)}({3.58*\j+2.19},.05);
  }
  \node at (0,-1.52) {\(\{a,c\}\)};
  \node at (3.58,-1.52) {\(\{c,u\}\)};
  \node at (7.16,-1.52) {\(\{u,d\}\)};
  \node at (10.74,-1.52) {\(\{d,b\}\)};
  \draw[thin,blue!45!black] (3.04,-1.84)--(3.04,-1.98)
    --(7.70,-1.98)--(7.70,-1.84);
  \node[below] at (5.37,-2.03) {\(\ell_1\leq h(Q)\leq\ell_2\)};
\end{tikzpicture}
\caption{The four configurations of the five-vertex graph. Filled
vertices are occupied. Each arrow is one reversible token slide.
Both middle configurations occupy \(u\).}\label{fig:clause-states}
\end{figure}

Take five vertices \(a,c,u,d,b\). Join every pair except
\(ac,cu,ud,db\). Figure~\ref{fig:clause-states} shows all six edges
and all four possible placements of the two tokens.

\begin{lemma}\label{lem:clause-states}
Under either move rule, the five-vertex graph has exactly four two-token
configurations. The possible moves between them are precisely
\[
\{a,c\}\longleftrightarrow\{c,u\}
\longleftrightarrow\{u,d\}\longleftrightarrow\{d,b\}.
\]
In particular, every reconfiguration from \(\{a,c\}\) to \(\{d,b\}\)
must occupy \(u\).
\end{lemma}

\begin{proof}
The only nonadjacent vertex pairs are \(ac,cu,ud,db\), so these are
exactly the legal two-token configurations. None admits a third vertex
nonadjacent to both occupied vertices, so these pairs are also maximum
independent sets.

Consecutive configurations in the displayed sequence are connected by
sliding a token along \(au,cd,ub\), respectively. Nonconsecutive
configurations share no vertex, so transforming one directly into another
would require moving both tokens. Thus no other single-token moves are
possible under either rule.
\end{proof}

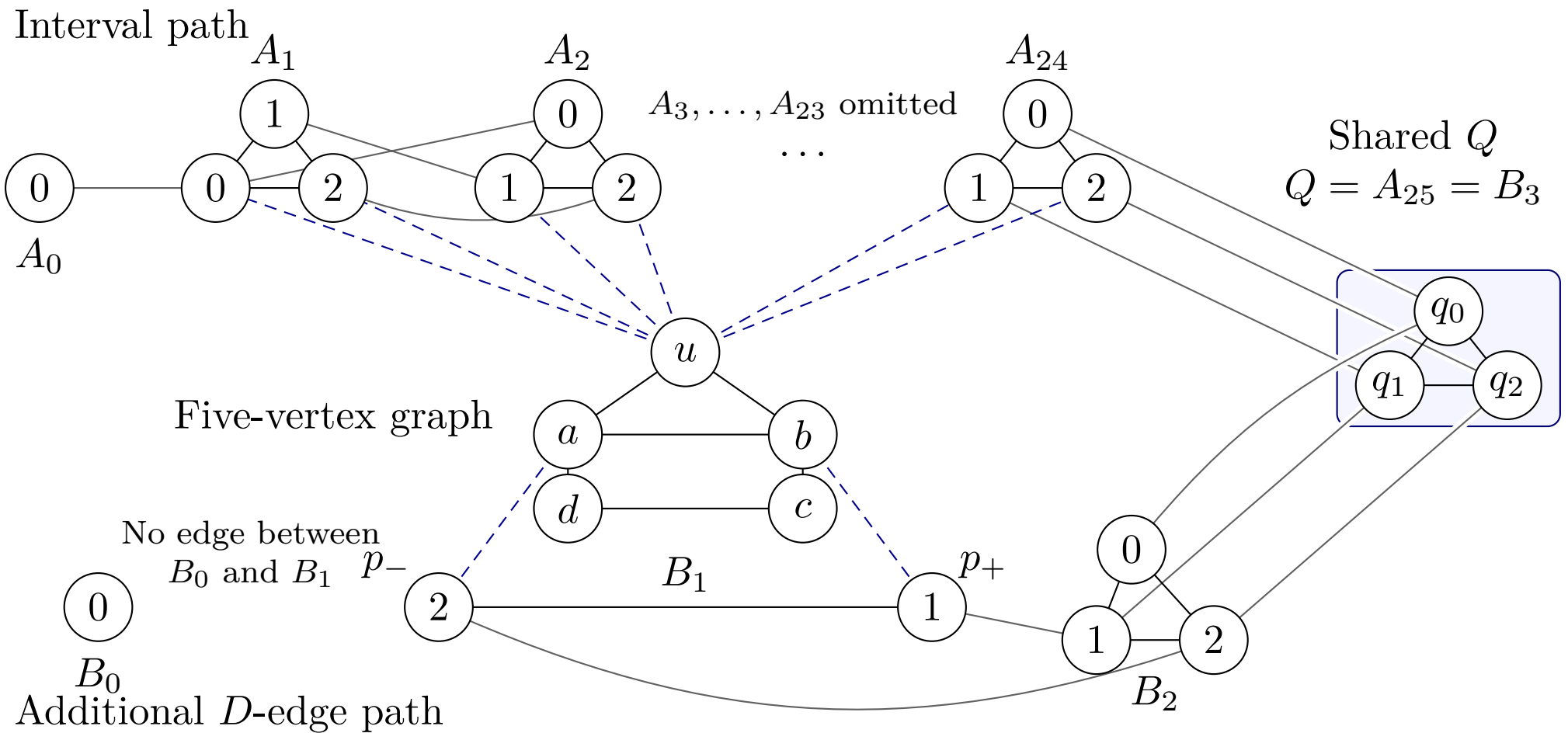
\begin{figure}[!ht]
\centering
\begin{tikzpicture}[x=1cm,y=0.7cm,font=\small,
 vertex/.style={circle,draw,fill=white,inner sep=0pt,minimum size=5.8mm},
 matching/.style={draw=black!62,preaction={draw=white,line width=2.3pt}},
 contact/.style={draw=blue!55!black,densely dashed}]
\coordinate (u) at (6,1.3);
\coordinate (a) at (5,.3);
\coordinate (b) at (7,.3);
\coordinate (d) at (5,-.6);
\coordinate (c) at (7,-.6);
\coordinate (B00) at (1,-1.8);
\coordinate (minus) at (3.9,-1.8);
\coordinate (plus) at (8.1,-1.8);
\coordinate (B20) at (9.8,-1.1);
\coordinate (B21) at (9.5,-2.2);
\coordinate (B22) at (10.5,-2.2);
\coordinate (q0) at (12.5,1.8);
\coordinate (q1) at (12, .9);
\coordinate (q2) at (13, .9);
\filldraw[fill=blue!4,draw=blue!45!black,rounded corners=3pt]
 (11.55,.4) rectangle (13.45,2.3);
\coordinate (A00) at (.5,3.3);
\coordinate (A11) at (2.5,4.2);
\coordinate (A10) at (2,3.3);
\coordinate (A12) at (3,3.3);
\foreach \j/\x in {2/5,24/9}{
 \coordinate (A\j0) at (\x,4.2);
 \coordinate (A\j1) at (\x-.5,3.3);
 \coordinate (A\j2) at (\x+.5,3.3);
}
\draw[matching] (A00)--(A10);
\draw[matching] (A10)--(A20) (A11)--(A21);
\draw[matching] (A12) to[bend right=22] (A22);
\foreach \i in {0,1,2}{\draw[matching] (A24\i)--(q\i);}
\draw[matching] (plus)--(B21);
\draw[matching] (minus) to[bend right=22] (B22);
\draw[matching] (B20) to[bend left=14] (q0);
\foreach \i in {1,2}{\draw[matching] (B2\i)--(q\i);}
\foreach \v in {A10,A12,A21,A22,A241,A242}{\draw[contact] (u)--(\v);}
\draw[contact] (a)--(minus) (b)--(plus);
\draw (a)--(u)--(b)--(c)--(d)--cycle (a)--(b);
\foreach \j in {1,2,24}{\draw (A\j0)--(A\j1)--(A\j2)--cycle;}
\draw (plus)--(minus) (B20)--(B21)--(B22)--cycle;
\draw (q0)--(q1)--(q2)--cycle;
\foreach \v in {a,b,c,d,u}{\node[vertex] at (\v) {\(\v\)};}
\foreach \i in {0,1,2}{\node[vertex] at (q\i) {\(q_\i\)};}
\node[vertex] at (A00) {0};
\foreach \j in {1,2,24}{\foreach \i in {0,1,2}{
 \node[vertex] at (A\j\i) {\(\i\)};
}}
\node[vertex] at (B00) {0};
\node[vertex] at (plus) {1};
\node[vertex] at (minus) {2};
\foreach \i in {0,1,2}{\node[vertex] at (B2\i) {\(\i\)};}
\node[anchor=west] at (.15,5.25) {Interval path};
\node at (7,3.7) {\(\cdots\)};
\node[font=\scriptsize] at (7,4.3) {\(A_3,\ldots,A_{23}\) omitted};
\node[below=3mm] at (A00) {\(A_0\)};
\foreach \j/\x in {1/2.5,2/5,24/9}{\node at (\x,4.95) {\(A_{\j}\)};}
\node[anchor=east] at (4.5,.5) {Five-vertex graph};
\node[align=center] at (12.2,3.6) {Shared \(Q\)\\\(Q=A_{25}=B_3\)};
\node[anchor=west] at (.15,-3.1) {Additional \(D\)-edge path};
\node[below=3mm] at (B00) {\(B_0\)};
\node[above left=1mm] at (minus) {\(p_-\)};
\node[above right=1mm] at (plus) {\(p_+\)};
\node at (6,-1.4) {\(B_1\)};
\node at (10,-2.85) {\(B_2\)};
\node[font=\scriptsize,align=center] at (2.3,-1.15)
 {No edge between\\\(B_0\) and \(B_1\)};
\end{tikzpicture}

\smallskip
\begin{tabular}{c@{\qquad}c@{\qquad}c}
Height & Vertices forced unoccupied & Forced clause pair\\
\hline
\(h(Q)=0\) & \(b\) (first sign \(+\)), \(u\) & \(\{a,c\}\)\\
\(h(Q)=D\) & \(a\) (first sign \(-\)), \(u\) & \(\{d,b\}\)
\end{tabular}
\caption{The clause attachments for \(D=3\) and
\([\ell_1,\ell_2]=[1,2]\). The interval path has \(m=12\) and
\(25\) column steps; the additional path has three. Both end at the
\emph{same} triangle \(Q\). Dashed blue edges are the clause contacts:
\(ap_-\), \(bp_+\), and the edges from \(u\) to the two
nonprescribed vertices of each displayed interval triangle. The same
rule applies to the omitted columns \(A_3,\ldots,A_{23}\). Solid
edges within columns are black; matching edges between columns are gray.
Crossings introduce no vertices.
The drawing specifies the graph, not a token configuration. The table
summarizes the forced states proved in Lemma~\ref{lem:clause-interval}.}
\label{fig:clause-attachments}
\end{figure}

\paragraph{Attaching the clause to the shared triangle.}
Let \(Q=\{q_0,q_1,q_2\}\) be the shared triangle introduced in
Section~\ref{sec:variables}, initially occupied at \(q_0\). We attach
\emph{two separate paths of columns} to this same triangle. The first
controls when \(u\) may be occupied; the second forces the clause to
start at \(\{a,c\}\) and finish at \(\{d,b\}\) as \(h(Q)\) goes from
zero to \(D\). Figure~\ref{fig:clause-attachments} shows the attachments
in a concrete example.

\emph{The interval path.} Apply Section~\ref{sec:interval} with its
triangle \(T\) identified with \(Q\), its outside vertex identified
with the existing clause vertex \(u\), and interval
\([\ell_1,\ell_2]\). Thus, for \(m=3(D+1)\), add a path
\(A_0,\ldots,A_N\), where \(N=2m+(\ell_2-\ell_1)\) and \(A_N=Q\).
All columns except \(Q\) are new. The fixed singleton \(A_0\) has
index \(-\ell_1-\ell_2\) modulo three. In each of
\(A_0,\ldots,A_{2m}\), join \(u\) to every vertex except the one
selected by the prescribed alternating signs \(+,-,\ldots,+,-\).
Use the initial configuration from Section~\ref{sec:interval}.
Lemma~\ref{lem:interval} then restricts occupancy of \(u\) to
\(\ell_1\leq h(Q)\leq\ell_2\). If \(\ell_1=\ell_2\), then
\(A_{2m}=Q\), so the same attachment rule also applies within \(Q\).

\emph{The additional \(D\)-edge path.} Add a second path
\(B_0,B_1,\ldots,B_D\), with \(B_D=Q\) and all other columns new.
The column \(B_0\) is a fixed singleton of index zero. In \(B_1\),
retain only the vertices of indices one and two, denoted \(p_+\) and
\(p_-\), respectively. Occupying \(p_+\) gives a plus as the first
sign; occupying \(p_-\) gives a minus. Add exactly the two edges
\[
                         bp_+,\qquad ap_-.
\]
Thus \(p_+\) blocks \(b\), whereas \(p_-\) blocks \(a\).
The remaining internal columns are triangles. Between consecutive
columns, join vertices of the same index as usual; in particular,
there is no edge between \(B_0\) and \(B_1\), since they have no
index in common. Initially all signs are plus, so the token in \(Q\)
is at index zero because \(3\mid D\).

Each column holds one token. The two paths share only \(Q\), and their
only edges to the five-vertex graph are the specified contacts with
\(u\), \(a\), and \(b\). Each clause adds \(O(D)\) vertices and edges.

\begin{lemma}\label{lem:clause-interval}
The clause gadget has the following properties:
\begin{enumerate}
\item At \(h(Q)=0\), the two tokens in the five-vertex graph occupy
\(\{a,c\}\).
\item At \(h(Q)=D\), these tokens occupy \(\{d,b\}\).
\item A move changing the occupied pair in the five-vertex graph is
possible only when \(\ell_1\leq h(Q)\leq\ell_2\).
\end{enumerate}
\end{lemma}

\begin{proof}
Let \(\mu\) count the minus signs on the additional \(D\)-edge path.
Its left endpoint is fixed, and initially \(\mu_0=0\), so
Lemma~\ref{lem:column-path} gives \(\mu=h(Q)\). Hence
\(0\leq h(Q)\leq D\).

\emph{Height zero.} All signs are plus, so \(p_+\) is occupied and
its neighbor \(b\) is unoccupied. Also, \(u\) cannot be occupied by
Lemma~\ref{lem:interval}, since \(\ell_1\geq1\). Among the four
pairs in Lemma~\ref{lem:clause-states}, only \(\{a,c\}\) avoids both
\(b\) and \(u\).

\emph{Height \(D\).} All signs are minus, so \(p_-\) is occupied
and its neighbor \(a\) is unoccupied. Again \(u\) cannot be occupied,
since \(\ell_2\leq D-1\). The only pair avoiding both \(a\) and
\(u\) is \(\{d,b\}\).

\emph{Changes of the clause pair.} By Lemma~\ref{lem:clause-states},
every move between the four pairs has \(u\) occupied immediately
before or after it. Lemma~\ref{lem:interval} therefore requires
\(\ell_1\leq h(Q)\leq\ell_2\). The move takes place within the
five-vertex graph, so \(Q\) stays fixed and has the same height on
both sides of the move.
\end{proof}

\begin{lemma}\label{lem:clause-switch}
For every integer \(t\in[\ell_1,\ell_2]\), the clause gadget admits
a reconfiguration from its initial configuration in which \(Q\) advances
from height \(0\) to \(D\) using only forward slides. The clause pair
changes from \(\{a,c\}\) to \(\{d,b\}\) at height \(t\), with \(Q\)
fixed during the change.
\end{lemma}
\begin{proof}
The key is to move the token in \(B_1\) at the right moment. Initially,
the clause occupies \(a\), so this token cannot occupy \(p_-\) and
must occupy \(p_+\). However, \(p_+\) blocks \(b\), which the clause
needs to occupy in its final pair. We therefore first move the clause
tokens to \(\{u,d\}\), leaving \(a\) and \(b\) empty. We can then
move the token from \(p_+\) to \(p_-\), freeing \(b\), and complete
the clause change.

We now show how to carry out this plan at the chosen height \(t\).

At height \(h=h(Q)\), the additional path has \(h\) minus signs and
\(D-h\) plus signs. Occupying \(p_+\) makes the first sign plus;
occupying \(p_-\) makes it minus. By Lemma~\ref{lem:column-path},
applied to \(B_1,\ldots,B_D\), the signs after the first position
can be rearranged while keeping \(B_1\) and \(Q\) fixed.

\emph{Reaching height \(t\).} Keep the clause at \(\{a,c\}\) and
the token in \(B_1\) at \(p_+\). Thus, the first sign remains plus.
Whenever \(h<t\), at least two plus signs remain, because
\(h<t\leq D-1\). One stays in the first position; bring another to
the last position to permit a forward slide of \(Q\). Since \(u\)
is empty, the interval gadget can also be prepared to permit this slide.
Repeat until \(h(Q)=t\).

\emph{Changing the clause at height \(t\).} Keep \(Q\) fixed. Since
\(t\geq1\), a minus sign exists. Bring it to the second position,
so that the first two signs are \(+,-\). Since
\(t\in[\ell_1,\ell_2]\), the interval gadget can free all its
neighbors of \(u\). Perform
\[
\{a,c\}\longrightarrow\{c,u\}\longrightarrow\{u,d\}.
\]

Now \(a\) and \(b\) are empty. The first two signs \(+,-\) give
occupied indices \(0,1,0\) in \(B_0,B_1,B_2\). Hence the token in
\(B_1\) can slide from \(p_+\) to \(p_-\): its destination has
index two, while the neighboring path tokens have index zero, and its
clause neighbor \(a\) is empty. This slide exchanges the first two signs,
\[
(+,-)\longrightarrow(-,+),
\]
and frees \(b\). Complete the clause change by sliding
\[
\{u,d\}\longrightarrow\{d,b\}.
\]

\emph{Continuing to height \(D\).} Keep the clause at \(\{d,b\}\)
and the token in \(B_1\) at \(p_-\), leaving \(b\) unblocked.
The first sign remains minus. Whenever \(h(Q)<D\), a plus sign
remains elsewhere in the path. Bring it to the last position, prepare
the interval gadget with \(u\) empty, and advance \(Q\). Repeating
this reaches height \(D\).
\end{proof}

In the full construction, the shared edge \(rw\) holds one token,
initially at \(w\), and \(w\) is adjacent to \(u\) in every clause.
Thus every change of the clause's occupied pair requires \(r\) to be
occupied.

\paragraph{The final copy.}
Take one last copy of the five-vertex graph and attach the interval
gadget for \([D,D]\) to its vertex \(u\), with the path ending at
\(Q\). This copy receives neither the additional \(D\)-edge path nor
the edge \(wu\). Changing from \(\{a,c\}\) to \(\{d,b\}\)
forces a visit to \(h(Q)=D\). At that height Lemma~\ref{lem:interval}
frees the path neighbors of \(u\), allowing the three slides in
Figure~\ref{fig:clause-states}. Both endpoint pairs leave \(u\) empty,
and their occupied vertices have no outside neighbors. Exchanging these
pairs therefore does not affect compatibility with the rest of the graph.

\section{The construction from 3-CNF-SAT}\label{sec:construction}

Let \(\Phi\) be a 3-CNF formula. Each clause contains at most three
literal occurrences, and repetitions are allowed. Discard variables that do not occur.
Number the remaining literal occurrences \(1,\ldots,s\) in clause
order. Give occurrence \(p\) the height \(6p\), and set \(D=6(s+1)\).
Thus all occurrence heights lie strictly between \(0\) and \(D\).

\paragraph{How the construction works.}
The construction can be understood as a sequence of literal
checks controlled by a single clock. The shared triangle \(Q\)
plays the role of this clock: its height \(h(Q)\) measures
progress through the formula, and occurrence \(p\) is assigned
the checking height \(6p\). Since height records net forward
progress rather than just the occupied vertex, the same
triangle can distinguish arbitrarily many checking times.

Each variable \(x\) has its own triangle \(V_x\), which
follows the clock either at the same height or one complete
turn ahead at checking times. These two possibilities
represent false and true, respectively. The variable gadget
allows either choice to be made while \(h(Q)=0\), but
prevents switching between them during a passage in which
\(h(Q)\) remains positive. Thus all checks made during
such a passage use one consistent assignment.

The edge \(rw\) acts as a shared gate for checking clauses.
Normally its token occupies \(w\), which is adjacent to
the vertex \(u\) of every ordinary clause gadget and
prevents any clause from changing its occupied pair.
To check a clause, the token must temporarily move from
\(w\) to \(r\). This move is possible only when every
gadget attached to \(r\) permits its occupancy.

The attachments ensure that a clause can be checked only
at appropriate times. Its interval gadget restricts the
clock to the interval containing its occurrences, while
the gadgets between consecutive occurrences eliminate
the intermediate checking heights. At the remaining
heights \(6p\), the occurrence gadget for literal \(p\)
forbids occupying \(r\) precisely when that literal is
false.

For example, consider the clause \(C=(x\lor\neg y)\),
with occurrences \(1\) and \(2\). Its interval is
\([6,12]\), and the gadget between the occurrences
excludes height \(9\). Since occupying \(r\) also
requires \(h(Q)\) to be a multiple of three, the clause
can be checked only at heights \(6\) and \(12\).
At height \(6\), the occurrence gadget for \(x\)
excludes \(h(V_x)=6\), which represents \(x\) being
false. The check is therefore possible only when
\(x\) is true, in which case \(h(V_x)=9\).
Similarly, at height \(12\), the occurrence gadget
for \(\neg y\) excludes \(h(V_y)=15\), which
represents \(y\) being true. This check is therefore
possible only when \(y\) is false, in which case
\(h(V_y)=12\). Thus the clause can change its
occupied pair only by checking a true literal.
The same mechanism applies to every clause, regardless
of the number of occurrences.

Finally, the starting and target configurations differ
only in a special final copy, which can change its occupied
pair only when \(h(Q)=D\). Reaching this height forces
every ordinary clause to change from its initial pair
to its final pair during the same positive passage.
Each clause must therefore have found a true literal
under the same assignment. Conversely, a satisfying
assignment allows us to advance the clock, check one
true literal per clause, and change the final copy.
Reversing the preceding moves then restores every
other token to its initial position and reaches the
target configuration.

\subsection{The graph}\label{sec:construction-graph}

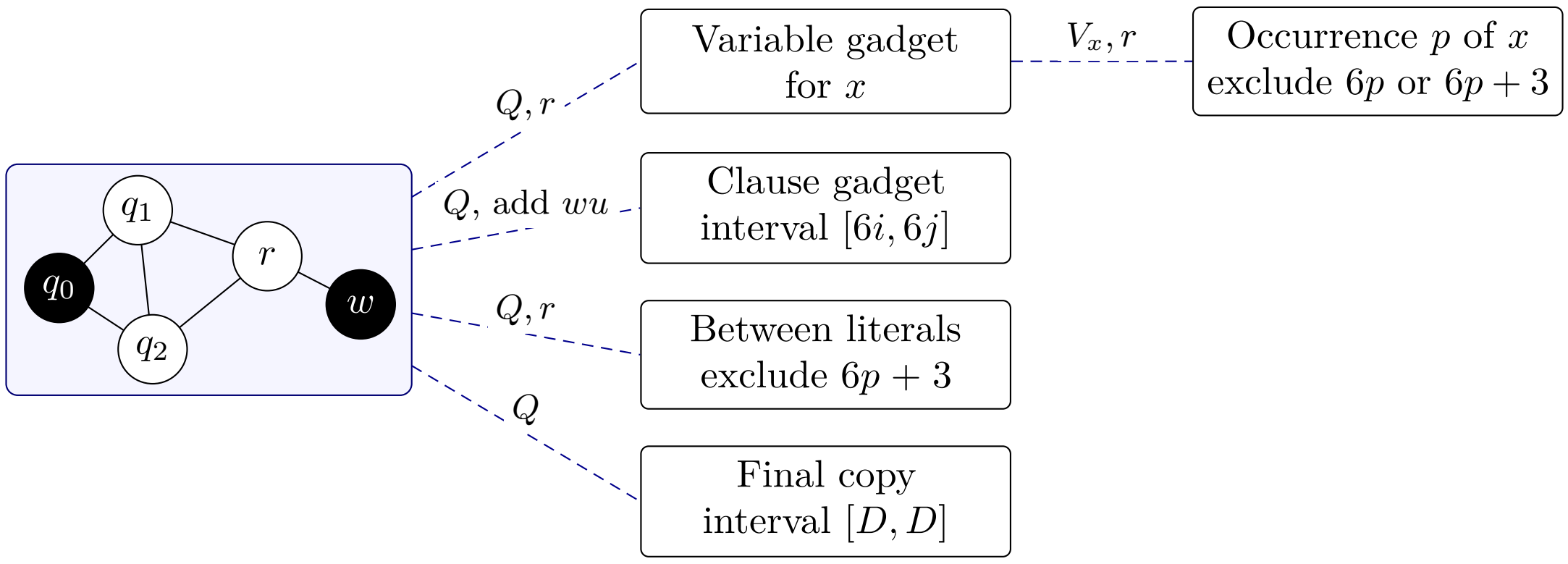
\begin{figure}[!ht]
\centering
\begin{tikzpicture}[
  x=.97cm,y=1cm,font=\small,
  vertex/.style={circle,draw,fill=white,inner sep=0pt,minimum size=6.3mm},
  token/.style={vertex,fill=black,text=white},
  gadget/.style={draw,rounded corners=2pt,align=center,fill=white,
                 text width=3.12cm,minimum height=9.5mm},
  attachment/.style={draw=blue!55!black,densely dashed},
  shared/.style={fill=white,inner sep=2pt,font=\footnotesize}]
  \filldraw[fill=blue!4,draw=blue!45!black,rounded corners=3pt]
    (-1.88,-1.04) rectangle (1.94,1.07);
  \coordinate (q0) at (-1.38,-.06);
  \coordinate (q1) at (-.64,.65);
  \coordinate (q2) at (-.50,-.62);
  \coordinate (r) at (.58,.23);
  \coordinate (w) at (1.46,-.21);
  \draw (q0)--(q1)--(q2)--cycle (r)--(q1) (r)--(q2) (r)--(w);
  \foreach \v/\name/\sty in {q0/q_0/token,q1/q_1/vertex,q2/q_2/vertex,
                             r/r/vertex,w/w/token}{
    \node[\sty] at (\v) {\(\name\)};
  }
  \node[gadget] (var) at (5.84,2.01) {Variable gadget\\for \(x\)};
  \node[gadget] (clause) at (5.84,.67) {Clause gadget\\interval \([6i,6j]\)};
  \node[gadget] (between) at (5.84,-.67) {Between literals\\exclude \(6p+3\)};
  \node[gadget] (last) at (5.84,-2.01) {Final copy\\interval \([D,D]\)};
  \node[gadget] (literal) at (11.04,2.01) {Occurrence \(p\) of \(x\)\\exclude \(6p\) or \(6p+3\)};
  \draw[attachment] (1.94,.77)--node[shared,above] {\(Q,r\)} (var.west);
  \draw[attachment] (1.94,.29)--node[shared,above] {\(Q\), add \(wu\)} (clause.west);
  \draw[attachment] (1.94,-.29)--node[shared,above] {\(Q,r\)} (between.west);
  \draw[attachment] (1.94,-.77)--node[shared,above] {\(Q\)} (last.west);
  \draw[attachment] (var.east)--node[shared,above] {\(V_x,r\)} (literal.west);
\end{tikzpicture}
\caption{Assembly of the construction. Solid edges form the shared core,
with initial tokens at \(q_0,w\). Each dashed connection represents an
attachment: its label gives the shared vertices, except that an ordinary
clause uses the common \(Q\) and receives an additional edge \(wu\).
All other vertices are new except where identifications are indicated.}
\label{fig:construction}
\end{figure}

We now assemble the gadgets using a shared clock and a shared gate.
Start with the triangle \(Q=\{q_0,q_1,q_2\}\), the edge \(rw\),
and the edges \(rq_1,rq_2\). Every gadget referring to \(Q\) or \(r\)
uses these same vertices. For each variable \(x\), all gadgets referring
to \(V_x\) likewise use one common variable triangle, identified
vertex by vertex according to its indices. Apart from these
identifications and the edges specified below, vertices added by
different gadgets are distinct. Figure~\ref{fig:construction}
summarizes the attachments.

\paragraph{Variables and clauses.}
For every variable \(x\), attach the variable gadget of
Section~\ref{sec:variables}, using the shared \(Q,r\) and a new
triangle \(V_x\). For each clause \(C\), let \(i\) and \(j\) be
the indices of its first and last literal occurrences. Because
occurrences are numbered in clause order, those of \(C\) are
exactly \(i,i+1,\ldots,j\). Attach the clause gadget of
Section~\ref{sec:clauses} to \(Q\) with interval \([6i,6j]\).
Write \(u_C\) for its vertex \(u\), and add the edge \(wu_C\).
As long as the token on \(rw\) occupies \(w\), all vertices
\(u_C\) are blocked. Since changing a clause's occupied pair
requires occupying \(u_C\), every such change requires the token
on \(rw\) to occupy \(r\).

Also attach the final copy from Section~\ref{sec:clauses}. Its
five-vertex graph has an interval gadget for \([D,D]\) at its
vertex \(u\), with the path ending at \(Q\). Unlike an ordinary
clause, this copy has neither an additional \(D\)-edge path
nor an edge from \(w\) to \(u\).

\paragraph{Literal occurrences.}
For each occurrence \(p\) of a variable \(x\), attach a fresh
excluded-height gadget from Section~\ref{sec:exclude}, with
\(T=V_x\), the common vertex \(r\), and \(B=D+3\).
Choose the height excluded when \(r\) is occupied as
\[
 t=\begin{cases}
     6p   & \text{if the literal is }x,\\
     6p+3 & \text{if the literal is }\neg x.
   \end{cases}
\]
This excludes precisely the height that would make the literal
false. Indeed, suppose \(h(Q)=6p\) and \(r\) is occupied during
the positive passage of Lemma~\ref{lem:variable-fixed}.
The edges from \(r\) force the tokens of \(Q\) and \(V_x\)
to occupy index zero. Consequently, Lemma~\ref{lem:variable-fixed}
gives \(h(V_x)=6p\) when \(x\) is false and
\(h(V_x)=6p+3\) when \(x\) is true. In either case,
Lemma~\ref{lem:exclude-height} forbids \(r\) at this
checkpoint if the literal is false. When the literal is true, this occurrence gadget can be prepared
to allow \(r\) along the forward sequence of
Lemma~\ref{lem:exclude-slides}.
Both choices of \(t\) are positive multiples of three below
\(B\). Moreover, the pairs \(\{6p,6p+3\}\) for distinct
occurrences \(p\) are disjoint, even when they belong to the
same variable.

\paragraph{Between consecutive occurrences.}
For each pair of consecutive occurrences \(p,p+1\) belonging to
the same clause, attach a fresh excluded-height gadget with
\(T=Q\), the common \(r\), \(B=D\), and \(t=6p+3\).
Whenever \(r\) is occupied, its edges to \(q_1,q_2\) force
\(Q\) to index zero, so \(h(Q)\) must be a multiple of three.
The height \(6p+3\) is the only such height strictly between
\(6p\) and \(6(p+1)\). Thus, for a clause with interval
\([6i,6j]\), its interval gadget and these gap exclusions leave
only the candidate checking heights
\(6i,6(i+1),\ldots,6j\). The occurrence gadgets then test
whether the corresponding literals are true. No gap gadget is
needed for a clause with just one occurrence.

\paragraph{Size of the construction.}
There are at most \(s\) variable gadgets, at most \(s\) ordinary
clause gadgets, and at most \(2s\) excluded-height gadgets:
one per occurrence and one per gap between consecutive occurrences
of a clause. Each gadget, including the final copy, has
\(O(D)\) vertices and edges. Because \(D=6(s+1)=O(s)\),
the graph has \(O(s^2)\) vertices and edges and can be
constructed in polynomial time.

\paragraph{Worked example.}
Consider \(\Phi=(x\lor\neg y)\land(\neg x\lor y)\), whose
four occurrences, in order, are \(x,\neg y,\neg x,y\).
Here \(D=30\). The clause intervals are \([6,12]\) and
\([18,24]\), and the two gap gadgets exclude heights \(9\)
and \(21\) on \(Q\). The occurrence gadgets exclude
\(h(V_x)=6\), \(h(V_y)=15\), \(h(V_x)=21\), and
\(h(V_y)=24\), respectively.

\begin{figure}[!t]
\centering
\begin{tikzpicture}[
    x=1cm,y=1cm,
    font=\footnotesize,
    module/.style={draw=black!55,rounded corners=2pt,fill=black!2,
                   line width=.45pt},
    important/.style={draw=blue!60!black,rounded corners=2pt,
                      fill=blue!5,line width=.65pt},
    thinrule/.style={black!35,line width=.45pt},
    yes/.style={circle,draw=blue!65!black,fill=blue!9,
                minimum size=5.2mm,inner sep=0pt},
    no/.style={circle,draw=black!55,fill=black!6,
               minimum size=5.2mm,inner sep=0pt}]

  \node[anchor=west,font=\bfseries\small] at (.20,10.60)
       {(a) The whole instance at gadget level};
  \node[anchor=east] at (15.1,10.60)
       {$\Phi=(x\lor\neg y)\land(\neg x\lor y)$, \quad $D=30$};

  \draw[important] (.20,9.28) rectangle (15.10,10.30);
  \node[anchor=west,align=left] at (.54,9.79)
       {\textbf{Shared clock} $Q$ \quad $h(Q):0\to30$};
  \node[anchor=west] at (7.90,9.79) {\textbf{Shared gate:}};
  \node[circle,draw,minimum size=5.6mm,inner sep=0pt] (r) at (11.07,9.79) {$r$};
  \node[circle,draw,fill=black,text=white,minimum size=5.6mm,
        inner sep=0pt] (w) at (12.40,9.79) {$w$};
  \draw[thick] (r)--(w);
  \node[anchor=west,font=\scriptsize] at (12.80,9.79) {initial token};

  \draw[module] (.20,7.77) rectangle (4.66,9.08);
  \node[align=center] at (2.43,8.42)
       {\textbf{Variable gadgets} $(Q,r)$\\
        $V_x,\ V_y$\\[-1pt]
        \scriptsize encode the two truth values};

  \draw[module] (4.83,7.77) rectangle (11.43,9.08);
  \node[align=center] at (8.13,8.68)
       {\textbf{Occurrence exclusions} $(V_x\text{ or }V_y,r)$};
  \node[align=center,font=\scriptsize] at (8.13,8.12)
       {$p_1:x\ [V_x,6]\qquad p_2:\neg y\ [V_y,15]$\\
        $p_3:\neg x\ [V_x,21]\qquad p_4:y\ [V_y,24]$};

  \draw[module] (11.60,7.77) rectangle (15.10,9.08);
  \node[align=center] at (13.35,8.42)
       {\textbf{Gap exclusions} $(Q,r)$\\
        $h(Q)=9,\ 21$\\[-1pt]
        \scriptsize forbid in-between checks};

  \draw[module] (.20,6.38) rectangle (5.01,7.53);
  \node[align=center] at (2.60,6.95)
       {\textbf{Clause $C_1=x\lor\neg y$}\\
        interval $[6,12]$; uses $Q$, edge $wu_1$};
  \draw[module] (5.19,6.38) rectangle (10.00,7.53);
  \node[align=center] at (7.60,6.95)
       {\textbf{Clause $C_2=\neg x\lor y$}\\
        interval $[18,24]$; uses $Q$, edge $wu_2$};
  \draw[module] (10.18,6.38) rectangle (15.10,7.53);
  \node[align=center] at (12.64,6.95)
       {\textbf{Final copy} (uses $Q$)\\
        interval $[30,30]$; no edge to $w$};

  \draw[thinrule] (.20,6.14)--(15.10,6.14);
  \node[anchor=west,font=\bfseries\small] at (.20,5.86)
       {(b) One successful forward passage};
  \node[anchor=east] at (15.1,5.86)
       {$x=y=\mathrm{true}$: $h(V_x)=h(V_y)=h(Q)+3$ at checks};

  \def\xzero{1.45}
  \def\xsix{4.06}
  \def\xnine{5.365}
  \def\xtwelve{6.67}
  \def\xeighteen{9.28}
  \def\xtwentyone{10.585}
  \def\xtwentyfour{11.89}
  \def\xthirty{14.50}

  \draw[blue!65!black,line width=2pt] (\xsix,5.15)--(\xtwelve,5.15);
  \draw[blue!65!black,line width=2pt] (\xeighteen,5.15)--(\xtwentyfour,5.15);
  \node[font=\scriptsize,above] at (5.365,5.19) {$C_1$ can change only in $[6,12]$};
  \node[font=\scriptsize,above] at (10.585,5.19) {$C_2$ can change only in $[18,24]$};

  \draw[->,line width=.9pt] (1.12,4.43)--(14.83,4.43);
  \node[anchor=east] at (1.00,4.43) {$h(Q)$};
  \foreach \xx/\lab in {\xzero/0,\xsix/6,\xnine/9,\xtwelve/12,
                          \xeighteen/18,\xtwentyone/21,
                          \xtwentyfour/24,\xthirty/30}{
       \draw[black!65] (\xx,4.34)--(\xx,4.52);
       \node[below=3pt,font=\scriptsize] at (\xx,4.33) {$\lab$};
  }
  \node[yes] at (\xsix,4.82) {$\checkmark$};
  \node[no] at (\xnine,4.82) {$\times$};
  \node[no] at (\xtwelve,4.82) {$\times$};
  \node[no] at (\xeighteen,4.82) {$\times$};
  \node[no] at (\xtwentyone,4.82) {$\times$};
  \node[yes] at (\xtwentyfour,4.82) {$\checkmark$};
  \node[font=\scriptsize,above] at (\xthirty,4.62) {final flip};

  \foreach \xx in {\xsix,\xtwelve,\xeighteen,\xtwentyfour}{
       \draw[thinrule] (\xx,4.00)--(\xx,3.70);
  }
  \draw[important] (2.90,2.41) rectangle (5.20,3.70);
  \node[align=center,font=\scriptsize] at (4.05,3.05)
       {$p_1:x$\\ $h(V_x)=9\ne 6$\\ \textbf{check $C_1$}};
  \draw[module] (5.52,2.41) rectangle (7.82,3.70);
  \node[align=center,font=\scriptsize] at (6.67,3.05)
       {$p_2:\neg y$\\ $h(V_y)=15=15$\\ blocked};
  \draw[module] (8.13,2.41) rectangle (10.43,3.70);
  \node[align=center,font=\scriptsize] at (9.28,3.05)
       {$p_3:\neg x$\\ $h(V_x)=21=21$\\ blocked};
  \draw[important] (10.74,2.41) rectangle (13.04,3.70);
  \node[align=center,font=\scriptsize] at (11.89,3.05)
       {$p_4:y$\\ $h(V_y)=27\ne24$\\ \textbf{check $C_2$}};

  \node[font=\scriptsize,anchor=west] at (.35,1.90)
       {Clause state changes (each passes through its vertex $u$):};
  \draw[important] (.40,.68) rectangle (5.05,1.61);
  \node[align=center] at (2.72,1.15)
       {$C_1:\ \{a,c\}\xrightarrow{\ h(Q)=6\ }\{d,b\}$};
  \draw[important] (5.25,.68) rectangle (9.95,1.61);
  \node[align=center] at (7.60,1.15)
       {$C_2:\ \{a,c\}\xrightarrow{\ h(Q)=24\ }\{d,b\}$};
  \draw[important] (10.15,.68) rectangle (15.05,1.61);
  \node[align=center] at (12.60,1.15)
       {$\mathrm{final}:\ \{a,c\}\xrightarrow{\ h(Q)=30\ }\{d,b\}$};
  \node[align=center,font=\scriptsize] at (7.65,.24)
       {After the final flip, reverse the outward moves; only the final pair remains changed.};
\end{tikzpicture}
\caption{A worked reduction for
$\Phi=(x\lor\neg y)\land(\neg x\lor y)$, with literal occurrences
$x,\neg y,\neg x,y$ numbered $1,2,3,4$ and $D=30$.
Panel (a) suppresses the internal vertices of the gadgets; labels in
parentheses specify shared attachment vertices, while $[V,t]$ records
an occurrence exclusion at $h(V)=t$. The two ordinary clauses also
have edges $wu_1$ and $wu_2$. This is an assembly schematic, not an
adjacency drawing. Panel (b) shows the forward passage for
$x=y=\mathrm{true}$. The gaps $9,21$ are excluded; at $12,18$,
the false literals block $r$. The two clauses change at heights $6,24$,
respectively, before the final copy changes at $30$.}
\label{fig:worked-reduction}
\end{figure}
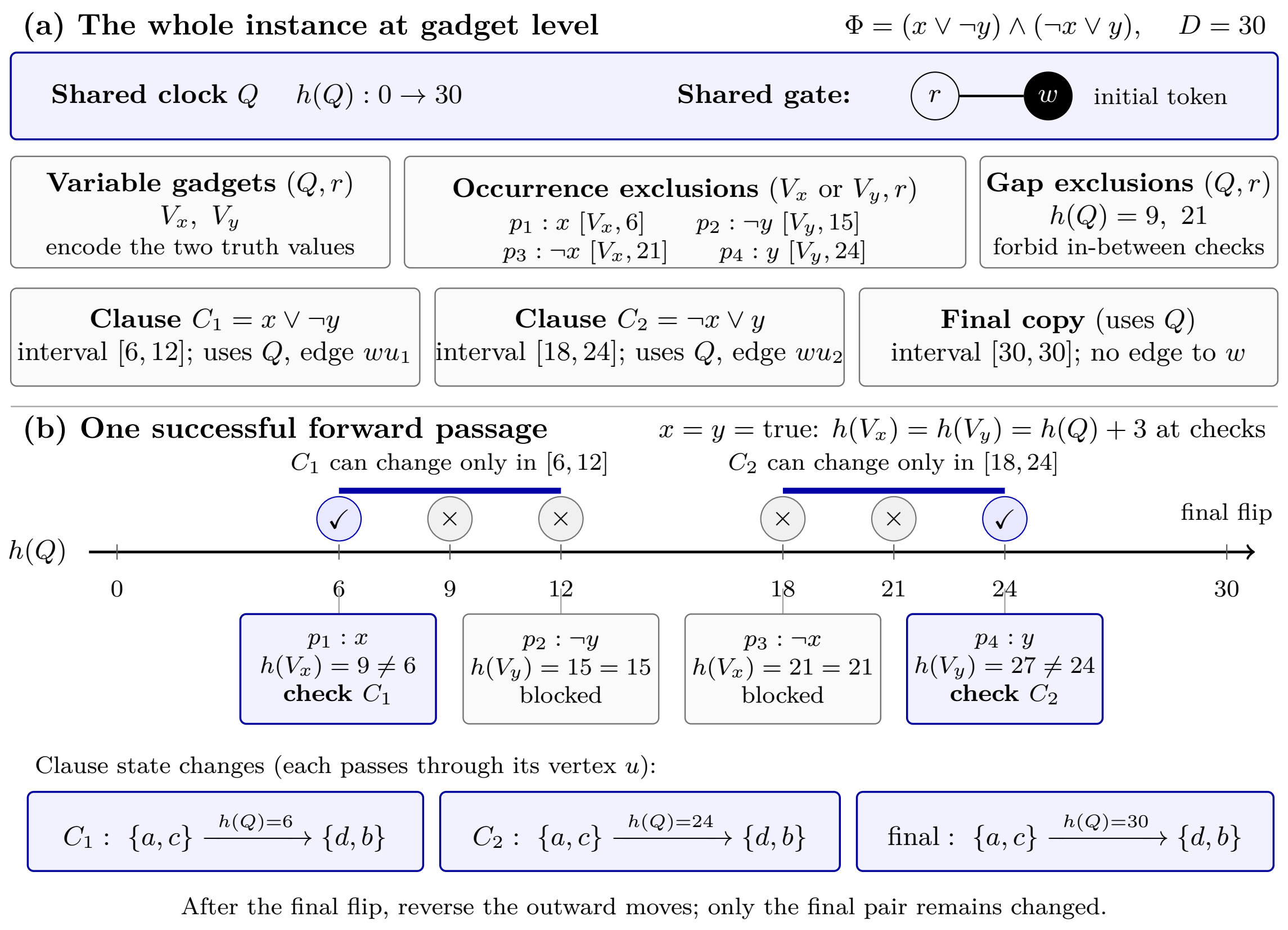

Take the satisfying assignment \(x=y=\mathrm{true}\).
Along the forward passage, at every checking height \(6p\)
we have \(h(V_x)=h(V_y)=6p+3\). At height \(6\), the
first literal \(x\) can be checked: \(h(V_x)=9\) avoids
its excluded height \(6\). At height \(24\), the last
literal \(y\) can likewise be checked, since
\(h(V_y)=27\) avoids \(24\). These checks change the
occupied pairs of the two clauses. By contrast, the potential
checks at \(12\) and \(18\) are blocked: their false literals
\(\neg y\) and \(\neg x\) yield the excluded heights
\(h(V_y)=15\) and \(h(V_x)=21\), respectively. The gap
heights \(9\) and \(21\) cannot be used for checks either.
After both clauses have changed, the clock reaches \(D=30\),
the final copy changes, and the earlier moves are reversed.
Figure~\ref{fig:worked-reduction} shows the gadget-level assembly
and this forward passage.

\subsection{The initial and target configurations}\label{sec:input-sets}

Define \(I_{\mathrm{start}}\) by the following placements.
\begin{itemize}
\item Occupy \(q_0\) in \(Q\) and \(w\) on \(rw\). In every variable
gadget occupy \(v_0\) in \(V_x\) and \(b\) in its triangle
\(\{a,b,z\}\). In each clause copy and in the final copy occupy
\(\{a,c\}\) in the five-vertex graph.
\item In each chain occupy its left singleton and \(p_0\) in \(P\).
Each of its triangles \(T_0,\ldots,T_m\) is at index one. In each
five-cycle occupy \(R_0=\{g_4,g_0\}\), and in its two auxiliary
triangles occupy \(a_1,b_1\), the vertices adjacent to \(r\).
\item Use the initial signs of each path. The variable path of length
\(D+3\) and the clause path of length \(D\) start with all signs plus.
For an interval \([\ell_1,\ell_2]\), start at index
\(-\ell_1-\ell_2\) with \(3(D+1)-\ell_1\) minus signs first, as in
Section~\ref{sec:interval}. On each path from \(P\) to the triangle
whose height is tested, start at index zero with \(L-1-t\) minus
signs first, where \(L\) and \(t\) are as in Section~\ref{sec:exclude}.
The remaining signs are plus. Starting at the left endpoint and
following the signs determines every occupied vertex on each path.
\end{itemize}

To obtain \(I_{\mathrm{goal}}\), replace \(\{a,c\}\) by \(\{d,b\}\)
in the final five-vertex graph and keep every other token in place.
Both configurations are determined by the formula alone.

\begin{lemma}\label{lem:maximum-endpoints}
\(I_{\mathrm{start}}\) and \(I_{\mathrm{goal}}\) are maximum independent
sets of the same size \(k\).
\end{lemma}
\begin{proof}
We first verify that both endpoint configurations are independent
sets, and then prove that they have maximum cardinality.

\smallskip
\noindent\emph{Independence of \(I_{\mathrm{start}}\).}
We check that the initial token placements are compatible with all
edges of the construction.

Consider first the paths of columns. Each column initially contains
exactly one token. By the definition of the signs, consecutive
columns have different occupied indices, so no matching-index edge
between consecutive columns has both endpoints occupied. We also
verify that the prescribed initial signs agree with the occupied
vertices of the triangles at which the paths terminate.

The height-bounding paths in the variable gadgets have length
\(D+3\), and the additional paths in the clause gadgets have length
\(D\). Both start at a singleton of index zero and initially have
only plus signs. Since \(D\) is divisible by three, both path
lengths are multiples of three, and their terminal triangles
therefore occupy index zero.

For an interval gadget with interval \([a,b]\), recall that
\(m=3(D+1)\). Its path has length \(2m+(b-a)\), starts at
index \(-a-b\), and initially contains \(m-a\) minus signs.
The net index change along this path is
\[
\begin{aligned}
\#(+)-\#(-)
&=2m+(b-a)-2(m-a)\\
&=a+b.
\end{aligned}
\]
Thus its terminal triangle also occupies index zero.

Similarly, in an excluded-height gadget, the path from \(P\)
to the tested triangle \(T\) has length \(L-2\) and initially
contains \(L-1-t\) minus signs. Its net index change is
\[
(L-2)-2(L-1-t)=2t-L,
\]
which is divisible by three. Since \(P\) initially occupies
\(p_0\), the token in \(T\) occupies index zero, as required.
Consequently, all paths are compatible with their prescribed
initial endpoint configurations.

Next consider the chains of five-cycle gadgets. Every triangle
\(T_0,\ldots,T_m\) initially occupies index one, and every
private five-cycle occupies the independent pair
\(R_0=\{g_4,g_0\}\). By the compatibility table in
Section~3.2, \(R_0\) permits the indices \((1,1)\) in its
two terminal triangles. Consecutive copies therefore agree
on their shared triangle. At the left endpoint, the occupied
singleton \(d\) is adjacent only to the index-zero vertex
of \(T_0\), which is unoccupied. At the right endpoint,
the tokens of \(T_m\) and \(P\) occupy indices one and zero,
respectively, so their matching-index edges cause no conflict.
Furthermore, the two auxiliary triangles of each copy initially
occupy \(a_1\) and \(b_1\). Each of these vertices has only
\(r\) as a neighbor outside its own triangle, and \(r\)
is initially unoccupied. Hence all initial placements within
the chains are compatible.

In every variable gadget, the initial occupied vertices are
\(q_0\), \(v_0\), and \(b\), where \(b\) belongs to the
auxiliary triangle \(\{a,b,z\}\). These placements are
compatible with the edges \(q_1v_0\), \(q_2v_1\),
\(aq_0\), and \(bv_2\): neither \(q_1\), \(q_2\),
\(a\), nor \(v_2\) is occupied. Moreover, \(z\) is
unoccupied, so the interval gadget attached to \(z\)
creates no conflict. The vertex \(r\) is also unoccupied,
so its edges to the variable triangles cause no conflict.

In every ordinary clause gadget, the five-vertex graph initially
occupies \(\{a,c\}\), which is an independent pair.
The additional path starts with a plus sign, so its first
two-vertex column occupies \(p_+\), while \(p_-\) is
unoccupied. The attachment edges \(bp_+\) and \(ap_-\)
therefore cause no conflict: \(b\) is unoccupied and
\(p_-\) is unoccupied. The interval gadget also causes no
conflict, since its external vertex \(u\) is initially
unoccupied. The same reasoning applies to the initial
pair \(\{a,c\}\) in the final five-vertex graph.

Finally, the shared edge \(rw\) initially has its token
on \(w\), while \(r\) is unoccupied. The vertex \(w\)
is adjacent to \(r\) and to the vertices \(u\) of the
ordinary clauses. All these neighbors are unoccupied.
Every remaining edge of the construction belongs to one
of the gadgets or attachments checked above. Therefore,
no edge has both endpoints occupied, and
\(I_{\mathrm{start}}\) is an independent set.

\smallskip
\noindent\emph{Independence of \(I_{\mathrm{goal}}\).}
By construction, \(I_{\mathrm{goal}}\) is obtained from
\(I_{\mathrm{start}}\) by replacing the occupied pair
\(\{a,c\}\) with \(\{d,b\}\) in the final five-vertex
graph, leaving every other token unchanged. Both pairs
are independent by Lemma~14. Moreover, the only vertex
of this final graph having neighbors outside it is \(u\),
which is unoccupied in both endpoint pairs. Thus the
replacement introduces no conflict with any token outside
the final graph. It follows that \(I_{\mathrm{goal}}\)
is also independent.

\smallskip
\noindent\emph{Maximum cardinality.}
Partition \(V(G)\) into the following pairwise disjoint parts:
all columns and auxiliary triangles, the edge \(rw\),
the private five-cycles, and the five-vertex graphs belonging
to the ordinary clauses and the final copy. Singleton
columns are included as columns, and every triangle shared
by several gadgets, in particular \(Q\) or some \(V_x\),
is included exactly once. All other vertices are private
to their respective parts, so this is a partition of \(V(G)\).

Write this partition as
\[
V(G)=P_1\mathbin{\dot\cup}\cdots
\mathbin{\dot\cup}P_N
\]
and let \(c_i=\alpha(G[P_i])\), the maximum size of an
independent set in the subgraph induced by \(P_i\).

Every column, auxiliary triangle, and the edge \(rw\)
is a nonempty clique, so its corresponding value \(c_i\)
equals one. Each private five-cycle has independence
number two. Each five-vertex graph also has independence
number two, since Lemma~14 identifies its four possible
independent pairs and shows that no independent set of
size three exists.

For any independent set \(I\) of \(G\), its intersection
with each part is independent in the corresponding
induced subgraph. Therefore,
\[
|I\cap P_i|\le c_i
\qquad\text{for every }i.
\]
Since the parts are disjoint and cover \(V(G)\),
\[
|I|
=\sum_{i=1}^{N}|I\cap P_i|
\le\sum_{i=1}^{N}c_i.
\]
Define
\[
k:=\sum_{i=1}^{N}c_i.
\]
Thus \(\alpha(G)\le k\).

By the initial placements, \(I_{\mathrm{start}}\)
contains exactly one token in every column and auxiliary
triangle, one token on \(rw\), two tokens in every private
five-cycle, and two tokens in every five-vertex graph.
Hence
\[
|I_{\mathrm{start}}\cap P_i|=c_i
\qquad\text{for every }i,
\]
and consequently \(|I_{\mathrm{start}}|=k\).

The target configuration changes only the occupied pair
in the final five-vertex graph, without changing its token
count. It therefore also satisfies
\[
|I_{\mathrm{goal}}\cap P_i|=c_i
\qquad\text{for every }i,
\]
giving \(|I_{\mathrm{goal}}|=k\).

Since both endpoint configurations are independent sets
of size \(k\), while no independent set can have more
than \(k\) vertices, we conclude that
\[
\alpha(G)
=|I_{\mathrm{start}}|
=|I_{\mathrm{goal}}|
=k.
\]
Thus both endpoint configurations are maximum independent
sets of the same size.
\end{proof}

The preceding proof also establishes the token-count
invariants needed for the gadget arguments. Indeed,
let \(I\) be any independent set of size \(k\). Since
\[
|I\cap P_i|\le c_i
\quad\text{for all }i,
\qquad
\sum_{i=1}^{N}|I\cap P_i|
=k=\sum_{i=1}^{N}c_i,
\]
equality must hold in every part:
\[
|I\cap P_i|=c_i
\qquad\text{for every }i.
\]
Otherwise, a deficit in one part would force another part
to exceed its capacity, which is impossible.

Now consider any legal reconfiguration move from such a
configuration. If a token moved from a part \(P_i\)
to a distinct part \(P_j\), the resulting configuration
would contain \(c_j+1\) tokens in \(P_j\), contradicting
\(\alpha(G[P_j])=c_j\). Therefore every legal move stays
within a single part. In particular, every column and
auxiliary triangle retains exactly one token, every
private five-cycle retains two tokens, and the edge
\(rw\) retains one token throughout any reconfiguration
sequence.

Finally, consider a legal token jump from a maximum
independent set \(I\), moving a token from \(u\in I\)
to an unoccupied vertex \(v\). If \(u\) and \(v\)
were nonadjacent, then the legality of the jump would
imply that \(v\) is nonadjacent to every vertex of
\(I\setminus\{u\}\). Together with the assumed
nonadjacency of \(u\) and \(v\), this would make
\(I\cup\{v\}\) an independent set of size \(k+1\),
contradicting \(\alpha(G)=k\). Hence every legal
token jump has adjacent endpoints and is therefore
a token slide.

The output instance is the graph \(G\), the token
count \(k\), and the configurations
\(I_{\mathrm{start}}\) and \(I_{\mathrm{goal}}\).

\section{From a satisfying assignment to a reconfiguration sequence}\label{sec:forward}

\begin{lemma}\label{lem:forward}
If \(\Phi\) is satisfiable, then \(I_{\mathrm{start}}\) reaches
\(I_{\mathrm{goal}}\) by token slides.
\end{lemma}
\begin{proof}
Fix a satisfying assignment, with \(\beta_x=0\) for false and
\(\beta_x=1\) for true. Choose one true occurrence in each clause.

\emph{Combining moves.}
Between checks, \(w\) is occupied and all clause vertices \(u\) are
empty. Before a slide in \(Q\) or \(V_x\), run each incident gadget up
to that slide, keeping the shared triangles fixed. Only private tokens
move during these preparations. By Figure~\ref{fig:construction}, no
edge joins private vertices in different gadgets. Once every incident
gadget permits the slide, make it once in the full graph.

\emph{Setting the variables and advancing \(Q\).}
Keep the token on \(w\) between checks. With \(h(Q)=0\), leave each
false variable at height zero. For each true variable, make one full
forward turn in \(V_x\) while its auxiliary token is on \(z\), then
return that token to \(b\). In each of \(D\) rounds, advance every
\(V_x\) once, one at a time, then advance \(Q\) once.
Lemma~\ref{lem:variable-choice} gives the auxiliary slides for this
order. After initialization and at the end of every round,
\begin{equation}\label{eq:forward-heights}
                       h(V_x)=h(Q)+3\beta_x
                       \qquad\text{for every variable }x .
\end{equation}

For each gadget of Section~\ref{sec:exclude}, follow the reconfiguration
sequence of Lemma~\ref{lem:exclude-slides} as its triangle \(Q\) or
\(V_x\) advances. A gadget on a true variable follows the three initial
slides as well. For a false variable stop at \(h(V_x)=D\).
Each clause follows Lemma~\ref{lem:clause-switch} with its chosen
height, keeping \(\{a,c\}\) before the check and \(\{d,b\}\) afterwards.
The final copy stays at \(\{a,c\}\), so Lemma~\ref{lem:interval}
permits the slides in \(Q\) with its vertex \(u\) empty.

\emph{Checking a clause.}
For a chosen occurrence \(p\), pause just after \(Q\) reaches height
\(6p\), before any variable advances again. By~\eqref{eq:forward-heights},
every \(V_x\) has height \(6p\) or \(6p+3\). Thus \(Q\) and all \(V_x\)
are at index zero, where they do not block \(r\).

The literal at occurrence \(p\) is true, so the height of its variable
differs from the excluded height. For any other
occurrence \(j\), the excluded height belongs to \(\{6j,6j+3\}\),
disjoint from \(\{6p,6p+3\}\). The gadgets attached to \(Q\) exclude
only heights congruent to three modulo six. Thus in every copy of
Section~\ref{sec:exclude} the current height \(h(T)\) and its excluded
height \(t\) are distinct multiples of three. If \(h(T)\leq t-3\), the
chain in Lemma~\ref{lem:exclude-slides} is still in its initial
configuration. If \(h(T)\geq t+3\), it has already reached its final
configuration. In both cases Lemma~\ref{lem:chain-slides} frees the
chain's neighbors of \(r\) using only its private auxiliary triangles.
Keep \(Q\) and all \(V_x\) fixed and make these slides in each copy in
turn. Every neighbor of \(r\) except \(w\) is now empty.

Following Lemma~\ref{lem:clause-switch}, bring a minus to the second
edge of the selected clause's \(D\)-edge path and free the
interval path neighbors of \(u\). Slide from \(w\) to \(r\). Keeping
\(Q\) fixed, change the clause through the pairs
\[
       \{a,c\}\ \longrightarrow\ \{c,u\}\
       \longrightarrow\ \{u,d\}\ \longrightarrow\ \{d,b\}.
\]
Swap the first two signs of the \(D\)-edge path while the pair is
\(\{u,d\}\). After the last slide every clause vertex \(u\) is empty.
Return the token from \(r\) to \(w\). Reverse the auxiliary slides
that freed \(r\) in each copy, then resume its local sequence. The
clause stays at \(\{d,b\}\) with its first sign minus.
Repeating this at the chosen occurrences changes each clause once.

\emph{Reaching the target.}
At \(h(Q)=D\), use Lemma~\ref{lem:interval} for \([D,D]\) to free
the path neighbors of \(u\) in the final copy, with \(Q\) fixed.
This copy has no edge \(wu\).
Leave the token on \(w\) and make the three slides in
Figure~\ref{fig:clause-states} in the final copy alone. Undo all
earlier slides in reverse order, starting with the preparation of
the final interval path. On the return the final pair is \(\{d,b\}\)
where it was \(\{a,c\}\) on the outward passage. Both pairs leave
\(u\) empty, and only \(u\) has outside neighbors. Every reversed
slide is therefore legal. Together they restore all other tokens to
their positions in \(I_{\mathrm{start}}\), giving exactly \(I_{\mathrm{goal}}\).
\end{proof}

\section{From a reconfiguration sequence to a satisfying assignment}\label{sec:backward}

\begin{lemma}\label{lem:backward}
If \(I_{\mathrm{start}}\) reaches \(I_{\mathrm{goal}}\) under either
move rule, then \(\Phi\) is satisfiable.
\end{lemma}
\begin{proof}
Fix any such reconfiguration sequence. By
Section~\ref{sec:input-sets}, every move is a slide within one part,
and each part keeps its initial number of tokens. All heights below
are counted from \(I_{\mathrm{start}}\).

\emph{A passage common to all clauses.}
The final five-vertex graph starts at \(\{a,c\}\) and ends at
\(\{d,b\}\). By Lemma~\ref{lem:clause-states}, a token must occupy
\(u\) somewhere in between. Its interval is \([D,D]\), so
Lemma~\ref{lem:interval} gives \(h(Q)=D\) at that point. End the
passage at the first visit to \(D\), and start it
at the last preceding visit to zero. Such a visit exists since \(Q\)
starts at height zero. Its height changes by at most one per slide,
so it is positive after the start of the passage. By
Lemma~\ref{lem:clause-interval}, every clause is at \(\{a,c\}\) at
the start and at \(\{d,b\}\) at the end. Thus every clause must occupy
its own vertex \(u\) during this same passage.

\emph{One assignment.}
For each variable \(x\), Lemma~\ref{lem:variable-fixed} puts
\(h(V_x)-h(Q)\) in one fixed pair throughout the positive part of the
passage. Assign \(x\) false (\(\beta_x=0\)) for the pair \(\{0,1\}\)
and true (\(\beta_x=1\)) for \(\{3,4\}\). A backward slide cannot
switch between the pairs: the difference changes by at most one per
slide, and the intervening value \(2\) is forbidden at positive
height by Lemma~\ref{lem:variable-residue}. Whenever \(r\) is occupied,
its edges to \(q_1,q_2,v_1,v_2\) force both triangles to index zero.
Their heights are multiples of
three, so at every such check on the positive part of the passage
\[
                        h(V_x)=h(Q)+3\beta_x .
\]

\emph{A true literal in each clause.}
Take a clause with occurrence indices \(i,\ldots,j\), and a
configuration on the passage in which its vertex \(u\) is occupied.
Its interval gadget forces \(6i\leq h(Q)\leq 6j\).
The edge \(uw\) forces \(w\) to be empty, so the token on \(rw\)
occupies \(r\). Hence \(h(Q)\) is a multiple of three. For each
\(i\leq p<j\), the gadget between occurrences \(p\) and \(p+1\)
rules out \(h(Q)=6p+3\) when \(r\) is occupied
(Lemma~\ref{lem:exclude-height}). Thus \(h(Q)\) is a multiple of six
in \([6i,6j]\), say \(6p\) for an occurrence \(p\) of this clause.

Let \(x\) be the variable of occurrence \(p\). At this check,
\(h(V_x)=6p+3\beta_x\). If the literal is \(x\) and is false, then
\(\beta_x=0\) and \(h(V_x)=6p\). If it is \(\neg x\) and is false,
then \(\beta_x=1\) and \(h(V_x)=6p+3\). In either case the gadget for
occurrence \(p\) excludes precisely this height on \(V_x\), so
Lemma~\ref{lem:exclude-height} forbids \(r\) from being occupied.
This contradicts the check. The literal is therefore true, and every
clause has a true literal under the same assignment.
\end{proof}

By Lemmas~\ref{lem:forward} and~\ref{lem:backward}, \(\Phi\) is
satisfiable if and only if \(I_{\mathrm{start}}\) reaches
\(I_{\mathrm{goal}}\), under both token sliding and token jumping.

\section{The tree decomposition}\label{sec:width}

We build a tree of bags with \(Q\cup\{r\}\) at its root. Whenever we
join two pieces, we specify a bag on each side containing every vertex
the two pieces share. Adding one tree edge between those bags gives a
tree. For each shared vertex, that edge joins its two subtrees of bags.

\paragraph{Paths of columns.}
Start with the first column as a bag. To pass from one triangle to the
next, replace the old vertex of each index by the new one, in the order
\(0,1,2\). For each replacement, first add the new vertex in one bag,
then remove the old vertex in the next. The bag between these steps
covers the matching edge. The end bags contain the full triangles, and
each vertex occurs in consecutive bags. For a column with missing
indices, omit those vertices from the same bags. This gives at most
four vertices per bag. We may therefore include one additional vertex
in every bag on the path.

In the interval gadget on \(T,u\), include \(u\) throughout its path of
columns. This covers every edge from \(u\) to a column and gives the
end bag \(T\cup\{u\}\), also when the interval consists of one height.
We use the same replacement rule for an auxiliary triangle from
Observation~\ref{obs:triangle}. At a bag containing its three outside
neighbors \(z_1,z_2,z_3\), attach a bag \(\{z_1,z_2,z_3\}\). Replace each
\(z_i\) by its neighbor in the triangle, adding before removing.
These bags cover the three added edges and the triangle.

\paragraph{The five-cycle and the chain.}
For the gadget of Figure~\ref{fig:five-cycle}, start a path at
\(X\cup\{r\}\), take the seven bags below in order, and end at
\(Y\cup\{r\}\). Each row gives the four vertices besides \(r\).
\begin{center}
\begin{tabular}{c@{\qquad}l@{\qquad}l}
Bag & Vertices besides \(r\) & Some edges covered\\
\hline
1 & \(x_0,x_1,x_2,g_3\) & \(x_1g_3\)\\
2 & \(x_0,x_2,g_3,g_0\) & \(x_2g_0,\ g_3g_0\)\\
3 & \(x_0,g_3,g_0,g_1\) & \(x_0g_1,\ g_3g_1\)\\
4 & \(g_3,g_0,g_1,y_1\) & \(g_1y_1\)\\
5 & \(y_1,g_0,g_1,g_4\) & \(g_1g_4\)\\
6 & \(y_1,g_0,g_4,y_2\) & \(y_1y_2\)\\
7 & \(y_1,y_2,g_4,y_0\) & \(g_4y_0\)
\end{tabular}
\end{center}
Attach the leaf \(\{r,g_0,g_4,y_2,g_2\}\) to bag~6. It covers all four
edges at \(g_2\), to \(g_0,g_4,y_2,r\). The table covers the other cycle
edges and the other edges to \(X,Y\). The end bags cover both triangles,
and bags~1 and~3 cover \(rx_0\) and \(rg_1\). Attach the auxiliary
triangle for \(r,g_0,y_2\) at bag~6 and the one for \(r,g_3,y_1\) at
bag~4. Each vertex occurs consecutively on the displayed path, and each
branch attaches at a bag containing all vertices it shares with the path.

Join consecutive five-cycle decompositions at their end bags containing
the shared triangle \(T_j\) and \(r\). At \(T_0\cup\{r\}\), attach a
leaf containing \(d\) and its sole neighbor in \(T_0\). At the other
end, include \(r\) in the column bags from \(T_m\) to \(P\). The bag
\(P\cup\{r\}\) covers \(rp_1,rp_2\). For a gadget excluding one height,
continue with \(r\) along the path from \(P\) to \(T\). Its last bag
\(T\cup\{r\}\) contains all vertices shared with the rest of the graph.

\paragraph{Variables and literal occurrences.}
For each \(x\), start a path at the root \(Q\cup\{r\}\). Keep \(r\) and
make the successive replacements
\[
       q_0\longrightarrow v_2,\qquad q_1\longrightarrow v_0,\qquad
       q_2\longrightarrow v_1,
\]
again adding before removing. The last two replacements cover the two
edges between \(Q\) and \(V_x\). The first gives a bag
\(Q\cup\{r,v_2\}\). Attach \(Q\cup\{z,v_2\}\) there, and to this new
bag attach the interval gadget for \([0,0]\) at its end bag
\(Q\cup\{z\}\). Also extend a separate branch from \(Q\cup\{z,v_2\}\)
with the two bags
\[
             \{z,q_0,v_2,a\},\qquad \{z,v_2,a,b\}.
\]
They cover \(aq_0,bv_2\) and the triangle \(\{a,b,z\}\).

The variable path of bags ends at \(V_x\cup\{r\}\), which covers
\(rv_1,rv_2\). Attach the path of length \(D+3\) there by its
\(V_x\) bag. Attach the gadget for each occurrence of \(x\) there by
its \(V_x\cup\{r\}\) bag. Each gadget excluding a height between
consecutive occurrences attaches directly to the root by its
\(Q\cup\{r\}\) bag.

\paragraph{Clauses and the final copy.}
Attach a single bag \(Q\cup\{r,w\}\) to the root. It covers \(rw\),
and the root covers \(rq_1,rq_2\). For each clause, attach a separate
bag \(Q\cup\{w,u\}\) to this bag, where \(u\) belongs to that clause.
This covers \(wu\) and keeps the bags containing \(w\) connected.
Attach the clause's interval gadget by its bag \(Q\cup\{u\}\).

From this last bag, keep \(u\) along the separate \(D\)-edge path of
columns, working from \(Q\) back to its two-vertex column. This gives
a bag containing \(u\) and the plus and minus vertices of that column.
Attach the initial singleton as a separate leaf bag. It has no neighbor
in the two-vertex column because neither index is zero. Now replace
the minus vertex by \(a\) and the plus vertex by \(b\), keeping \(u\)
and adding before removing. This covers both edges from the column to
the five-vertex graph. At the resulting bag \(\{u,a,b\}\), attach the
leaf \(\{a,c,u,d,b\}\) to cover every edge of that graph. For the final
copy, join its interval bag \(Q\cup\{u\}\) directly to the root and
attach its five-vertex bag there.

\begin{lemma}\label{lem:width}
The construction gives a tree decomposition of \(G\) with \(O(s^2)\)
bags, each of size at most five.
\end{lemma}

\begin{proof}
We verify that the bags constructed above form a tree
decomposition of \(G\), that every bag has size at
most five, and that the total number of bags is
\(O(s^2)\).

\smallskip
\noindent\emph{Vertex and edge coverage.}
Every vertex of \(G\) belongs to one of the
columns, auxiliary triangles, five-cycles,
five-vertex graphs, or the shared edge \(rw\).
The constructions above introduce a bag containing
every vertex of each such part.

Every edge internal to a column, triangle, five-cycle,
or five-vertex graph is covered by the corresponding
local decomposition. Matching-index edges between
consecutive columns are covered by the bags in which
the new vertex is added before the old vertex is
removed.

The remaining edges are the attachments between
these parts. The interval and excluded-height
decompositions retain their attachment vertices
throughout the appropriate paths. The seven-bag
construction and its branches cover all edges of
each five-cycle gadget. The variable and clause
constructions explicitly cover their attachment
edges, including those incident to \(Q\), \(V_x\),
\(r\), \(w\), and the clause vertices \(u\).
Thus every edge of \(G\) has both endpoints in
some bag.

\smallskip
\noindent\emph{Connectedness.}
Within every path of columns, a vertex appears
in consecutive bags: an old vertex remains present
until it is removed, while a new vertex remains
present from the moment it is introduced.
The same property holds for the replacement paths
in the variable and clause constructions.

In the five-cycle decomposition, every vertex
appears in consecutive bags along the main path.
Each additional branch is attached at a bag
containing all vertices it shares with that path.
Consequently, the bags containing any vertex form
a connected subtree within each gadget decomposition.

The full decomposition is obtained by joining
these local decomposition trees one at a time.
Each join uses a single tree edge between bags
containing every vertex shared by the two pieces.
Hence the resulting bag graph is a tree.
Moreover, for every shared vertex, this edge
connects the subtrees of bags containing that
vertex on the two sides. Thus the bags containing
each vertex of \(G\) form a connected subtree
of the full decomposition.

\smallskip
\noindent\emph{Bag size.}
During a column replacement, at most four
vertices are present in a bag. Retaining one
additional attachment vertex increases this
number to at most five. The seven main bags
of each five-cycle decomposition, its additional
leaf, and all explicitly listed attachment bags
also contain at most five vertices.
Therefore every bag has size at most five, and
the resulting tree decomposition has width
at most four.

\smallskip
\noindent\emph{Number of bags.}
Each column replacement and each five-cycle
contributes only a constant number of bags.
Consequently, each variable gadget, clause
gadget, and excluded-height gadget contributes
\(O(D)\) bags, as does the final copy.
There are \(O(s)\) such gadgets, and
\(D=6(s+1)=O(s)\). Hence the total number
of bags is
\[
O(sD)=O(s^2).
\]
All bags and their connections are explicitly
specified by the construction and can be
generated in polynomial time.
\end{proof}

\paragraph{Why the width is four.}
Every formula used in Section~\ref{sec:construction} has at least one
literal occurrence. The gadget for that occurrence contains at least
two consecutive five-cycles,
since \(B=D+3\) and the chain has \(L/3\geq 2\) copies.
Take the three vertices of their shared triangle as three singleton
sets. Take the second cycle as a fourth set and the first cycle together
with \(r\) as a fifth. These five sets are disjoint and connected.
Each cycle has an edge to each vertex of the shared triangle, and \(r\)
has an edge to each cycle. Thus every two sets are adjacent. Contracting
each set gives a \(K_5\) minor.

In any tree decomposition, the bags meeting a connected vertex set
form a subtree, since the subtrees for the ends of each edge intersect.
The five sets above give five pairwise intersecting subtrees. Root the
tree and, for each subtree, take its node nearest the root. This node
lies on the root path to every node in that subtree. Choose a deepest
of the five nearest nodes, say \(t\), from subtree \(S\). For any other
subtree \(S'\), take a node in \(S\cap S'\). On the path from the root
to that node, \(t\) lies between the nearest node of \(S'\) and the
intersection node, so \(t\) belongs to \(S'\). The bag at \(t\) meets
all five disjoint sets. It therefore has at least five vertices.
Together with Lemma~\ref{lem:width}, this gives treewidth four.

\begin{proof}[Proof of Theorem~\ref{thm:main}]
The construction is polynomial, and Lemma~\ref{lem:maximum-endpoints}
gives two maximum independent sets. By Lemmas~\ref{lem:forward}
and~\ref{lem:backward}, they are reachable from one another under
either rule exactly when \(\Phi\) is satisfiable. The graph has
treewidth four and we supply its decomposition by Lemma~\ref{lem:width}.
NP-hardness follows from that of \textsc{3-CNF-SAT}.
\end{proof}

\section{Conclusion and open problems}\label{sec:conclusion}

We proved that ISR is NP-hard on graphs of treewidth four
under both token sliding and token jumping.
The complexity of both versions remains open at treewidth two and
three. Bousquet conjectured that both admit polynomial-time algorithms
at width two~\cite[Conjecture~68]{bousquet2024configuration}.
At width three, is either version already NP-hard? The \(K_5\) minor
in Section~\ref{sec:width} shows that a reduction to width three would
have to change our chain of five-cycles.
Both versions are PSPACE-complete for some fixed treewidth
bound~\cite{wrochna2018bandwidth,vanderzanden2015constraint}.
Whether PSPACE-hardness (or membership in NP) holds at treewidth four remains open for
each rule.

\section*{Use of AI}

We used OpenAI's GPT-6 Astra to study ISR. 
Four was the smallest width at which we proved NP-hardness for
both token sliding and token jumping. We spent approximately 48 hours in conversations 
exploring gadgets for treewidth four and an additional 48 hours on treewidth two and three without resolving either case.
We also used the model to look for counterexamples, check the gadget
proofs, and draft and revise the text and figures. We take full
responsibility for all the content of this paper.

\bibliographystyle{plainurl}
\bibliography{references}
\end{document}